\documentclass[preprint,aps]{revtex4}

\usepackage{graphicx}% Include figure files
\usepackage{dcolumn}% Align table columns on decimal point
\usepackage{bm}% bold math
\usepackage{epsf}

\newcommand{\be}{\begin{eqnarray}}

\newcommand{\Eq}[1]{Eq.~(\ref{#1})}
\newcommand{\Eqs}[2]{Eqs.(\ref{#1},\ref{#2})}
\newcommand{\Eqss}[3]{Eqs.(\ref{#1},\ref{#2},\ref{#3})}

\newcommand{\ur}[1]{(\ref{#1})}
\newcommand{\urs}[2]{(\ref{#1},\ref{#2})}
\newcommand{\urss}[3]{(\ref{#1},\ref{#2},\ref{#3})}

\newcommand{\beq}{\begin{equation}}
\newcommand{\eeq}{\end{equation}}

\newcommand{\la}[1]{\label{#1}}
\newcommand{\bea}{\begin{eqnarray}}
\newcommand{\eea}{\end{eqnarray}}
\newcommand{\beqa}{\begin{eqnarray}}
\newcommand{\eeqa}{\end{eqnarray}}
\newcommand{\ba}{\begin{array}}
\newcommand{\ea}{\end{array}}

\newcommand{\half}{{\textstyle{\frac{1}{2}}}}

\newcommand{\n}{\nonumber}
\newcommand{\nn}{\nonumber}
\newcommand{\bydef}{\stackrel{d}{=}}

\newcommand{\Tr}{{\rm Tr}}

\newcommand{\eps}{\epsilon^{\kappa\lambda\mu\nu}}

\renewcommand{\(}{\left(}
\renewcommand{\)}{\right)}
\renewcommand{\[}{\left[}
\renewcommand{\]}{\right]}

\def\Dirac#1{#1\hskip-6pt/}

\def\appendix{\par
\setcounter{subsection}{0}
\setcounter{equation}{0}

\def\thesection{Appendix}
\def\theequation{\Alph{section}.\arabic{equation}}}

\begin{document}

\title{\bf Low-energy General Relativity with torsion: \\
a systematic derivative expansion}

\author{Dmitri Diakonov$^{1,2}$, Alexander G. Tumanov$^2$ and Alexey A. Vladimirov$^3$}

\affiliation{$^1$ Petersburg Nuclear Physics Institute, Gatchina 188300, St. Petersburg, Russia \\
$^2$ St. Petersburg Academic University, St. Petersburg 194021, Russia \\
$^3$ Ruhr-Universit\"at Bochum, Bochum D-44780, Germany}

\date{November 28, 2011}

\begin{abstract}
We attempt to build systematically the low-energy effective Lagrangian for the Einstein--Cartan formulation of gravity theory that generally includes the torsion field. We list all invariant action terms in certain given order; some of the invariants are new. We show that in the leading order the fermion action with torsion possesses additional $U(1)_{\rm L}\times U(1)_{\rm R}$ gauge symmetry, with 4+4 components of the torsion (out of the general 24) playing the role of Abelian gauge bosons. The bosonic action quadratic in torsion gives masses to those gauge bosons. Integrating out torsion one obtains a point-like 4-fermion action of a general form containing vector-vector, axial-vector and axial-axial interactions. We present a quantum field-theoretic method to average the 4-fermion interaction over the fermion medium, and perform the explicit averaging for free fermions with given chemical potential and temperature. The result is different from that following from the ``spin fluid'' approach used previously. On the whole, we arrive to rather pessimistic conclusions on the possibility to observe effects of the torsion-induced 4-fermion interaction, although under certain circumstances it may have cosmological consequences. \\



%(Physics and Astronomy Classification Scheme, http://www.aip.org/pacs/pacs.html/)
% 04.20.Cv 	Fundamental problems and general formalism
% 04.60.Bc 	Phenomenology of quantum gravity
% 98.80.Cq 	Particle-theory and field-theory models of the early Universe
% 04.50.Kd 	Modified theories of gravity
% 98.80.-k 	Cosmology
% 11.15.-q 	Gauge field theories

\end{abstract}

\maketitle

\section{Introduction}

Recently, there has been some renewed interest in torsion appearing in the Einstein--Cartan formulation of General Relativity~\cite{Hehl:1976kj,Sezgin:1979zf,Hehl:1994ue,Shapiro:2001rz,Hammond:2002rm,Khriplovich:2005jh},
and in physical effects it may imply, see {\it e.g.} Refs.~\cite{Freidel:2005sn,Mielke:2006zp,Zubkov:2010sx}.
A general drawback of these interesting studies was certain lack of systematics, in particular not all {\it a priori} possible invariants of a given order containing torsion were considered. The first aim of this paper is to fill in this gap and to introduce torsion including its interaction with fermions in a systematic way, in the spirit of the low-energy derivative expansion~\cite{arXiv:0908.1964}. In the lowest order of this expansion, torsion induces a local 4-fermion interaction~\cite{Hehl:1976kj} which may affect cosmological evolution at very early times and high matter
density~\cite{Nurgaliev:1983,Gasperini:1986mv,deBerredoPeixoto:2009zz,Poplawski:2010kb}, 
or be detectable from high-precision data at later times. Therefore, the second aim of this paper is to evaluate the contribution of the general torsion-induced 4-fermion interaction derived in the first part, to the stress-energy tensor for possible applications. We think that such evaluation in the past, based on ``spin fluid'' ideas, has been unsound.\\

General Relativity with fermions is a theory invariant under {\it i}) general coordinate transformations (diffeomorphisms) and {\it ii}) point-dependent (local) Lorentz transformations. The standard way one introduces fermions is via the Fock--Weyl action~\cite{Fock:1929vt,Weyl:1929fm}
\beq
S_{\rm f}=i\int d^4x\,\det(e)\,\frac{1}{2}\left(\overline{\Psi}\,e^{A\mu}\,\gamma_A\,
{\cal D}_\mu\Psi
-\overline{{\cal D}_\mu\Psi}\,e^{A\mu}\,\gamma_A\,\Psi\right),\qquad
{\cal D}_\mu=\partial_\mu +\frac{1}{8}\omega_\mu^{BC}[\gamma_B\gamma_C],
\la{ferm-gen}\eeq
where $\Psi$ is the 4-component fermion field assumed to be a world scalar, $\gamma_A$ are the four Dirac matrices, $\omega_\mu^{BC}$ is the gauge field of the local Lorentz group, called spin connection, and $e^{A\mu}$ is the contravariant (inverse) frame field. In fact there are other fermion actions invariant under {\it i} and {\it ii}, to be discussed below, but \Eq{ferm-gen} is generic.

To incorporate fermions, one needs, therefore, the gauge field $\omega_\mu$ and  the frame field $e_\mu$, which are {\it a priori} independent. Therefore, the bosonic part of the General Relativity action must be also constructed from these fields. We are thus bound to the Einstein--Cartan formulation of General Relativity.

In this formulation, the lowest-derivative terms in the bosonic part of the action invariant under {\it i} and {\it ii} are
\beq\la{bos-lowest}
S_{\rm b}=\frac{M_P^2}{16\pi}\int d^4x\,\left(-2\Lambda\,\det(e)
-\frac{1}{4}\,\eps\,\epsilon_{ABCD}\,{\cal F}^{AB}_{\kappa\lambda}\,e^C_\mu e^D_\nu
-\frac{\iota}{2}\,\eps\,{\cal F}^{AB}_{\kappa\lambda}\,e^A_\mu e^B_\nu +\ldots \right)
\eeq
where $D_\mu^{AB}=(\partial_\mu+\omega_\mu)^{AB}$ is the covariant derivative, and
${\cal F}_{\mu\nu}^{AB}=[D_\mu D_\nu]^{AB}$ is the curvature. The first term is the cosmological term,
$\det(e)=(1/4!)\eps\,\epsilon_{ABCD}\,e^A_\kappa e^B_\lambda e^C_\mu e^D_\nu=\sqrt{-g}$,
with $\Lambda=5.5\cdot 10^{-84}\,{\rm Gev}^2$, the second term is the would-be Einstein--Hilbert action with $M_P=1/\sqrt{G}=1.22\cdot 10^{19}\,{\rm GeV}$ being the Planck mass, and the third term is the $P$- and $T$-odd action first suggested in Refs.~\cite{Hojman:1980kv,Nelson:1980ph} with {\em iota} being so far a free
dimensionless parameter. In the context of canonical gravity the inverse of $\iota$ is sometimes called the Barbero--Immirzi parameter, and the third term in \ur{bos-lowest} called the ``Holst action''~\cite{Holst:1995pc} although in fact they were introduced much earlier~\cite{Hojman:1980kv,Nelson:1980ph}. In a quantum theory one writes the amplitude as $\exp(iS)$.

\Eq{bos-lowest} is quadratic in $\omega_\mu$, therefore the saddle-point integration in $\omega_\mu$ in the assumed path integral over $\omega_\mu,\;e_\mu$, is exact. We define the antisymmetric torsion tensor as
\beq
T^A_{\mu\nu}\bydef \frac{1}{2}\left((D_\mu e_\nu)^A-(D_\nu e_\mu)^A\right).
\la{torsion-def}\eeq
Under Lorentz transformations it behaves as a 4-vector, and under diffeomorphisms it behaves as a rank-2 tensor. In the absence of fermions, the saddle-point equation arising from the first variation of \Eq{bos-lowest} in 24 independent variables $\omega^{AB}_\mu$ results in 24 dynamical equations $T^A_{\mu\nu}=0$. This set of equations is linear in $\omega_\mu$, and its solution is the well known
\beq
\bar\omega^{AB}_\mu(e)=\frac{1}{2}e^{A\kappa}(\partial_\mu e^B_\kappa-\partial_\kappa e^B_\mu)
-\frac{1}{2}e^{B\kappa}(\partial_\mu e^A_\kappa-\partial_\kappa e^A_\mu)
-\frac{1}{2}e^{A\kappa}e^{B\lambda}e^C_\mu(\partial_\kappa e^C_\lambda-\partial_\lambda e^C_\kappa).
\la{omega-bar}\eeq
Quantities with a bar refer, here and below, to this zero-torsion case. [If two flat indices $A,B,C,...$ appear both as subscripts or both as superscripts we sum over them with Minkowski signature $\eta = (1,-1,-1,-1)$; $\epsilon^{0123}=1,\;\epsilon_{0123}=-1$.] It is well known that in the zero-torsion case the second term in \Eq{bos-lowest} reduces to the standard Einstein--Hilbert action $-\sqrt{-g}\bar R$ where $\bar R$ is the standard scalar curvature built from the Christoffel symbol $\bar\Gamma_{\kappa\lambda,\mu}
=\half(\partial_\kappa g_{\lambda\mu}+\partial_\lambda g_{\kappa\mu}-\partial_\mu g_{\kappa\lambda})$, and the third term reduces to $-\half \bar R_{\kappa\lambda,\mu\nu}\eps=0$.

With fermions switched in, the torsion is nonzero even at the saddle point since from varying \Eq{ferm-gen} one obtains $T\sim J$, where $J\sim\bar\Psi\Psi$ is a fermion bilinear current. Neither is it generally speaking zero, if terms with higher derivatives are added to the bosonic action, see Section VI.C.

Apart from full derivatives, the deviation of the second and third term in \Eq{bos-lowest} from its zero-torsion limit is ${\cal O}(T^2)$, and the deviation of the fermionic action \ur{ferm-gen} from its zero-torsion limit is ${\cal O}(T)$, see \Eqss{EC1}{EC2}{S-gamma-M} below. However, these terms are not the only ones that can be constructed from the requirements {\it i} and {\it ii}, and there are no {\it a priori} reasons why other terms should be ignored. 

The minimal actions \urs{ferm-gen}{bos-lowest} are at best the low-energy limit of an effective theory whose microscopic origin is still under debate, since the action is non-renormalizable and, worse, non-positive definite. Therefore, the best we can do in the absence of a well-defined quantum theory is to write down a derivative expansion for an effective low-energy action satisfying the requirements {\it i} and {\it ii} with arbitrary constants, to be in principle determined or at least restricted from observation or experiment. A future fundamental, microscopic theory of gravity will be able to fix those constants. Unless proved otherwise, the derivative expansion in the effective action is assumed to be a Taylor series in $\partial^2/M_P^2$ in the bosonic sector; in the fermionic sector odd powers of $\partial/M_P$ are also allowed.

For completeness, we include in our consideration invariants that are odd under $P,T$ inversion. Since these discrete symmetries are not preserved by weak interactions, and the effective low-energy gravity may imply integrating out high-momenta fermions, we do not see the principle why such invariants should be avoided.

\section{Order of magnitude analysis}

In the Einstein--Cartan formulation of gravity, which as we stress is unavoidable if
we wish to incorporate fermions, the frame field $e^A_\mu$ and the spin connection $\omega^{AB}_\mu$ are {\it a priori} independent variables. One expands the action in the derivatives of $e_\mu$ and $\omega_\mu$, preserving the diffeomorphism invariance and gauge invariance under local Lorentz transformations, the only two symmetries requested. The spin connection $\omega_\mu$ is a gauge field that transforms inhomogeneously under local Lorentz transformations, hence it can appear only inside covariant derivatives.

At the saddle point (in other wording from the equation of motion) the spin connection is expressed through the frame field, $\omega_\mu = \bar \omega_\mu \sim e^{-1}\partial\, e$, see \Eq{omega-bar}. Suppressing the indices and omitting the frame fields assumed to be of the order of unity, one can present symbolically the general $\omega_\mu$ as
\beq
\omega = \bar \omega + T,\qquad T\sim \partial\, e,
\la{omega-expansion}\eeq 
where $T$ is the torsion field \ur{torsion-def}; the precise relation is given by \Eq{omega-full} below. From the point of view of derivative counting, $T$, $\bar\omega$ and hence $\omega$ itself are all one-derivative quantities. From the point of view of the gauge group, $\bar\omega$ transforms inhomogeneously and hence must always come inside covariant derivatives, whereas $T$ transforms homogeneously and hence terms of the type $T^2$ and the like are allowed by gauge symmetry.

Let us classify the possible action terms. There is only one zero-derivative term, the invariant volume or the cosmological term. One-derivative terms are absent in the bosonic sector. They appear only in the fermionic sector from the Dirac--Fock--Weyl action which we fully analyze in Section IV: there is a unique term of the type $\bar\Psi(\partial+\bar\omega)\Psi$ and four terms of the type $\bar\Psi T\Psi$.

There are precisely seven two-derivative terms: two terms linear in the curvature 
${\cal F}$, presented in \Eq{bos-lowest}, and five terms quadratic in torsion, fully listed in Section VI.A. Since by definition $T\sim \nabla e$, all terms quadratic in torsion are quadratic in the derivatives. There are no other terms quadratic in the derivatives in the bosonic sector, and this is an exact statement of this paper.

Turning to four-derivative terms, there are in general terms of the type 
${\cal F}{\cal F}\sim R^2$, $T(\nabla{\cal F})$, $(\nabla T)^2$, 
$T^2(\nabla T)$, $T^2{\cal F}$ and $T^4$. 

Omitting the cosmological term we write down symbolically the effective Lagrangian as
\bea\la{eff-lagr}
{\cal L}_{\rm eff}&=& M_P^2\bar R+\bar\Psi\partial\Psi+\bar\Psi T\Psi + M_P^2T^2\\ 
\n
&&+\;\;\bar R^2 + T(\bar\nabla\bar R)+(\bar\nabla T)^2+ T^2(\bar\nabla T)+T^2\bar R +T^4 
+ {\cal O}(1/M_P^2)
\eea
where the Lagrangian for matter is represented by the fermionic source. The equations
of motion are obtained by varying the action \ur{eff-lagr} with respect to the fields involved. Assuming $\hbar=c=1$ and the dimensionless metric tensor $g_{\mu\nu}$, the dimensions of the quantities in \Eq{eff-lagr} are the usual $R\sim 1/{\rm cm}^2$,
$T\sim 1/{\rm cm}$, $\bar\Psi\Psi \sim 1/{\rm cm}^3$. In the leading order one gets from the first two terms the standard estimate for the curvature generated by matter,
\beq
\bar R\;\sim\; \frac{<\bar\Psi \partial\Psi>}{M_P^2}\; \sim\; \frac{p^4}{M_P^2},
\la{simR}\eeq
where $p$ is the characteristic momentum of matter, be it temperature, cubic root of density, or mass -- in fact the largest of these. Being substituted back into the action, this estimate shows that the first two terms of the action \ur{eff-lagr} are of the order of $p^4$, and that the $\bar R^2$ term is a tiny $p^8/M_P^4$ correction.

Excluding torsion $T$ from the 3$^{\rm d}$ and 4$^{\rm th}$ term in \ur{eff-lagr} gives
\bea\la{simT}
&& T\;\sim\; \frac{<\bar\Psi \Psi>}{M_P^2}\;\sim\; \frac{p^3}{M_P^2},\\
\n
&& M_P^2T^2\;\sim\; T <\bar\Psi \Psi>\; \sim\;
\frac{<(\bar\Psi \Psi)(\bar\Psi \Psi)>}{M_P^2}\;\sim\; \frac{p^6}{M_P^2}.
\eea  
This is the leading post-Einstein correction, and we analyze its most general structure and its effect in this paper. From the estimate \ur{simT} we see that other terms in \Eq{eff-lagr} give even smaller corrections:
\bea\la{simFF} 
&&{\cal F}^2\;\sim\; T(\nabla{\cal F})\; \sim\; (\nabla T)^2\;\sim \;\frac{p^8}{M_P^4},\\
\n
&&T^2(\nabla T)\;\sim\; T^2{\cal F}\;\sim\; \frac{p^{10}}{M_P^6},\qquad T^4\sim\frac{p^{12}}{M_P^8}.
\eea 
Nevertheless, we list for completeness all 10 possible terms of the type ${\cal F}^2$
in Section VI.B and all 4 possible terms of the type $T(\nabla{\cal F})$ in Section VI.C.

\section{General framework}

In this section we introduce the basic variables and make sign and normalization conventions. To simplify the algebra, we temporarily deal with the Euclidean signature where the Lorentz group $SO(4)$ acting on flat indices $A,B,C,...$ is locally isomorphic to the direct product $SU(2)_{\rm L}\times SU(2)_{\rm R}$. We return to Minkowski signature in the final results.

\subsection{$SU(2)_{\rm L}\times SU(2)_{\rm R}$ subgroups of the Lorentz group}

The 4-component Dirac bi-spinor field in the spinor basis is
\beq
\Psi = \left(\begin{array}{c} \psi^\alpha \\
\chi^{\dot\alpha} \end{array}\right),\qquad
\Psi^\dagger=\left(\psi^\dagger_\alpha,\chi^\dagger_{\dot\alpha}\right).
\la{bi-spinor}\eeq
The Lorentz $SU(2)_{\rm L}\times SU(2)_{\rm R}$ transformation rotates the 2-component Weyl spinors:
\bea\n
\psi^\alpha &\to & \left(U_L\right)^\alpha_\beta\psi^\beta,\qquad
\psi^\dagger_\alpha\to \psi^\dagger_\beta \left(U^\dagger_L\right)^\beta_\alpha,\qquad
U_L\in SU(2)_{\rm L},\\
\chi^{\dot\alpha}&\to & \left(U_R\right)^{\dot\alpha}_{\dot\beta}\chi^{\dot\beta},\qquad
\chi^\dagger_{\dot\alpha}\to \chi^\dagger_{\dot\beta} \left(U^\dagger_R\right)^{\dot\beta}_{\dot\alpha},\qquad
U_R\in SU(2)_{\rm R}.
\la{Weyl_transf}\eea
In the spinor basis the (Euclidean) Dirac matrices are
\beq
\gamma_A=\left(\begin{array}{cc}0&\sigma^-_A\\ \sigma^+_A & 0\end{array}\right),
\qquad \gamma_5=\left(\begin{array}{cc}{\bf 1}_2&0\\ 0 & -{\bf 1}_2\end{array}\right),\qquad
\sigma^{\pm}_A=({\bf 1}_2,\pm i\,\tau^i),\quad \left(\sigma^+_A\right)^\dagger=\sigma^-_A.
\la{Dirac-gamma}\eeq
where $\tau^i\,(i=1,2,3)$ are the three Pauli matrices. We introduce the commutators
\bea\n
\Sigma^{+-}_{AB}&\bydef&\frac{i}{2}\left(\sigma^+_A\sigma^-_B-\sigma^+_B\sigma^-_A\right)
=-\eta^i_{AB}\tau^i,\qquad \left(\Sigma^{+-}_{AB}\right)^\dagger=\Sigma^{+-}_{AB},\\
\Sigma^{-+}_{AB}&\bydef&\frac{i}{2}\left(\sigma^-_A\sigma^+_B-\sigma^-_B\sigma^+_A\right)
=-\overline\eta^i_{AB}\tau^i,\qquad \left(\Sigma^{-+}_{AB}\right)^\dagger=\Sigma^{-+}_{AB},
\la{Sigmapm}\eea
where $\eta,\bar\eta$ are 't Hooft symbols. They are projectors of $so(4)$ to the two $su(2)$ subalgebras.
The basic relations for 't Hooft symbols are
\bea\n
&&\eta^i_{AB}\eta^i_{CD}=\delta_{AC}\delta_{BD}-\delta_{AD}\delta_{BC}+\epsilon_{ABCD},\\
\n
&&\bar\eta^i_{AB}\bar\eta^i_{CD}=\delta_{AC}\delta_{BD}-\delta_{AD}\delta_{BC}
-\epsilon_{ABCD},\\
\la{tHooft}
&&\eta^i_{AB}\eta^j_{AB}=4\delta^{ij},\qquad \bar\eta^i_{AB}\bar\eta^j_{AB}=4\delta^{ij},
\qquad \eta^i_{AB}\bar\eta^j_{AB}=0.
\eea
With Euclidean signature, there is no distinction between upper and lower flat (capital Latin) indices,
in particular, $\epsilon^{1234}=\epsilon_{1234}=1$.

The following commutation relations are helpful and will be used below:
\bea\n
&&\Sigma^{+-}_{BC}\sigma^+_A-\sigma^+_A\Sigma^{-+}_{BC}
=-2i\,\left(\delta_{BA}\sigma^+_C-\delta_{CA}\sigma^+_B\right),\\
&&\Sigma^{-+}_{BC}\sigma^-_A-\sigma^-_A\Sigma^{+-}_{BC}
=-2i\,\left(\delta_{BA}\sigma^-_C-\delta_{CA}\sigma^-_B\right).
\la{rot}\eea

\subsection{Tetrad}

We introduce the frame field in the matrix form:
\beq
e^\pm_\mu\bydef e^A_\mu\sigma^\pm_A.
\la{ematr}\eeq
Under Lorentz transformations, the frame field transforms as
\beq
e^+_\mu\to U_R e^+_\mu U_L^\dagger,\qquad e^-_\mu\to U_L e^-_\mu U_R^\dagger.
\la{transfe}\eeq
Under the general differentiable change of the coordinate system $x^\mu\to x'^{\mu}(x)$
the frame field transforms as a world vector:
\beq
e^\pm_\mu(x)\to e^\pm_\lambda(x'(x))\; \frac{\partial x'^\lambda}{\partial x^\mu}.
\la{diff}\eeq

\subsection{Covariant derivatives and curvatures}

The requirement that the theory is invariant under {\em local} (point-dependent)
$SO(4)$ Lorentz transformation demands that a `compensating' gauge field $\omega^{AB}_\mu=-\omega^{BA}_\mu$,
called spin connection, must be introduced. With its help, one constructs the covariant derivatives,
\bea\la{covderAB}
D^{AB}_\mu &=& \partial_\mu \delta^{AB}+\omega^{AB}_\mu,\\
\la{covdermp}
\nabla^{-+}_\mu &=& \partial_\mu\,{\bf 1}_2-\frac{i}{4}\omega^{AB}_\mu\Sigma^{-+}_{AB}\quad
=\quad \partial_\mu\,{\bf 1}_2-iL^i_\mu\frac{\tau^i}{2}\quad = \quad \partial_\mu - i L_\mu,\\
\la{covderpm}
\nabla^{+-}_\mu &=& \partial_\mu\,{\bf 1}_2-\frac{i}{4}\omega^{AB}_\mu\Sigma^{+-}_{AB}\quad
=\quad \partial_\mu\,{\bf 1}_2-iR^i_\mu\frac{\tau^i}{2}\quad = \quad \partial_\mu - i R_\mu,
\eea
where $L(R)$ are left (right) connections,
\beq
\omega^{AB}_\mu=-\frac{1}{2}\,L^i_\mu\,\bar\eta^i_{AB}-\frac{1}{2}\,R^i_\mu\,\eta^i_{AB}.
\la{Omdec}\eeq
Inversely,
\beq
L^i_\mu=-\frac{1}{2}\bar\eta^i_{AB}\omega^{AB}_\mu,\qquad R^i_\mu=-\frac{1}{2}\eta^i_{AB}\omega^{AB}_\mu.
\la{pirho}\eeq
In matrix notations
\beq
L_\mu=L^i_\mu\frac{\tau^i}{2},\qquad R_\mu=R^i_\mu\frac{\tau^i}{2}.
\la{LRmatr}\eeq
Under gauge transformation \ur{transfe} the covariant derivatives transform as follows:
\beq
\nabla^{-+}\to U_L\nabla^{-+}U_L^\dagger,\qquad \nabla^{+-}\to U_R\nabla^{+-}U_R^\dagger,
\la{transfnabla}\eeq
corresponding to the usual gauge transformation of the connections,
\beq
L_\mu \to U_L(L_\mu+i\partial_\mu)U_L^\dagger,\qquad R_\mu \to U_R(R_\mu+i\partial_\mu)U_R^\dagger.
\la{transfLR}\eeq

The commutators of the covariant derivatives are curvatures:
\bea\la{curvAB}
[D_\mu D_\nu]^{AB} &=&{\cal F}^{AB}_{\mu\nu}=\partial_\mu \omega^{AB}_\nu-\partial_\nu \omega^{AB}_\mu
+\omega^{AC}_\mu\omega^{CB}_\nu-\omega^{AC}_\nu\omega^{CB}_\mu,\\
\la{curvmp}
i[\nabla^{-+}_\mu\,\nabla^{-+}_\nu] &=& F^{-+}_{\mu\nu}
= \left(\partial_\mu L_\nu -\partial_\nu L_\mu-i[L_\mu L_\nu]\right)
= F^i_{\mu\nu}(L)\frac{\tau^i}{2},\\
\la{curvpm}
i[\nabla^{+-}_\mu\,\nabla^{+-}_\nu] &=& F^{+-}_{\mu\nu}
= \left(\partial_\mu R_\nu -\partial_\nu R_\mu-i[R_\mu R_\nu]\right)
= F^i_{\mu\nu}(R)\frac{\tau^i}{2}.
\eea
The $SO(4)$ curvature is decomposed accordingly into two pieces transforming as
the $({\bf 3,1})\oplus({\bf 1,3})$ representation of the $SU(2)_{\rm L}\times SU(2)_{\rm R}$ group:
\beq
{\cal F}^{AB}_{\mu\nu}
= -\frac{1}{2}\,F^i_{\mu\nu}(L)\,\bar\eta^i_{AB}-\frac{1}{2}\,F^i_{\mu\nu}(R)\,\eta^i_{AB}
\la{fsSOdec}\eeq
where
\bea\n
F^i_{\mu\nu}(L)&=&\partial_\mu L^i_\nu-\partial_\nu L^i_\mu +\epsilon^{ijk}L^j_\mu L^k_\nu,\\
\la{fs}
F^i_{\mu\nu}(R)&=&\partial_\mu R^i_\nu-\partial_\nu R^i_\mu +\epsilon^{ijk}R^j_\mu R^k_\nu
\eea
are the usual Yang--Mills field strengths of the $SU(2)$ Yang--Mills potentials $L^i_\mu$
and $R^i_\mu$. These field strengths are projections of the full curvature:
\bea\n
F^{-+}_{\mu\nu}(L) &=& F^i_{\mu\nu}(L)\frac{\tau^i}{2}\quad =\quad\frac{1}{4}\Sigma^{-+}_{AB}\,{\cal F}^{AB}_{\mu\nu},\\
F^{+-}_{\mu\nu}(R) &=& F^i_{\mu\nu}(R)\frac{\tau^i}{2}\quad =\quad\frac{1}{4}\Sigma^{+-}_{AB}\,{\cal F}^{AB}_{\mu\nu}.
\la{Fpm}\eea

\subsection{Torsion}

The antisymmetric combinations
\bea\n
&&(\nabla^{-+}_\mu e^-_\nu - e^-_\nu\nabla^{+-}_\mu) - (\nabla^{-+}_\nu e^-_\mu - e^-_\mu\nabla^{+-}_\nu)
= \left(D_\mu^{AB} e_\nu^B-D_\nu^{AB} e_\mu^B\right)\sigma_A^- \; = \; 2\,T^A_{\mu\nu}\sigma_A^-,\\
&&(\nabla^{+-}_\mu e^+_\nu - e^+_\nu\nabla^{-+}_\mu) - (\nabla^{+-}_\nu e^+_\mu - e^+_\mu\nabla^{-+}_\nu)
= \left(D_\mu^{AB} e_\nu^B-D_\nu^{AB} e_\mu^B\right)\sigma_A^+ \; = \; 2\,T^A_{\mu\nu}\sigma_A^+,
\la{torsion}\eea
define the torsion field
$T^A_{\mu\nu}\bydef \half\left((D_\mu e_\nu)^A-(D_\nu e_\mu)^A\right)\equiv (D_{[\mu} e_{\nu]})^A$.
It is a 4-vector with respect to Lorentz transformations, and an antisymmetric rank-2 tensor with respect
to diffeomorphisms. Contracting it with $e^{A\alpha}$ one gets a rank-3 tensor
$T^{~~\lambda}_{\mu\nu}=-T^{~~\lambda}_{\nu\mu}=T^A_{\mu\nu}e^{A\lambda}$.

The torsion tensor $T_{\mu\nu}^{~~\lambda}$ has 24 independent components. It is convenient
to decompose $T^{~~\lambda}_{\mu\nu}$ into the totally antisymmetric part related to an axial vector
($a^\kappa$, 4 components), the trace part related to a vector ($v_\kappa$, 4 components), and the rest
16 components ($t_{\mu\nu}^{~~\lambda}$) subject to constraints~\cite{Tsamparlis:1981xm,Hehl:1994ue,Capozziello:2001mq}:
\bea
T_{\mu\nu}^{~~\lambda}=\frac{2}{3}\,\det(e)\,a^\kappa\,\epsilon_{\kappa\mu\nu\rho}\,g^{\rho\lambda}
+ \frac{2}{3}\,v_\kappa\,\delta^{\kappa}_{[\mu}\delta^\lambda_{\nu]}
+\frac{2}{3}t_{\mu\nu}^{~~\lambda},
\la{T-split}\eea
where, inversely,
$$
a^\kappa=\frac{\epsilon^{\kappa\alpha\beta\gamma}}{4\det(e)}\,T_{\alpha\beta,\gamma},\qquad
v_\kappa=T^{~~\mu}_{\kappa\mu},\qquad
t_{\mu\nu}^{~~\lambda} = T_{\mu\nu}^{~~\lambda} + \delta^\lambda_{[\mu}\,T^{~~\rho}_{\nu]\rho}
- g^{\lambda\rho}\,T_{\rho[\mu,\nu]}\,.
$$
The reduced torsion tensor $t^{~~\rho}_{\mu\nu}$ satisfies 4 constraints $\eps\,t_{\lambda\mu,\nu}=0$ and
4 constraints $t^{~~\rho}_{\mu\rho}=0$, therefore it has 16 degrees of freedom, as it should.

The general 24-component spin connection can be presented as a sum of the zero-torsion part \ur{omega-bar} and the torsion part:
\bea\la{omega-full}
\omega_\mu^{AB} &=& \bar\omega_\mu^{AB}
+ e^A_\alpha e^{B\beta}\(T_{~\mu\beta}^{\alpha}+T_{~\beta\mu}^{\alpha}+T_{\mu\beta}^{~~\alpha}\) \\
\n
&=& \bar\omega_\mu^{AB} + \frac{2}{3}\,\epsilon^{ABCD}e^C_\kappa e^D_\mu a^\kappa
+ \frac{4}{3}\, e^{[B}_\mu e^{A]\kappa} v_\kappa
+ \frac{2}{3}\,e^{A\alpha}e^{B\beta}\(t_{\alpha\mu,\beta}+t_{\alpha\beta,\mu}+t_{\mu\beta,\alpha}\).
\eea

\subsection{Affine connection}

In the Einstein--Cartan formulation, the tetrad $e^A_\mu$ and the spin connection
$\omega^{AB}_\mu$ are primary fields, whereas the metric tensor $g_{\mu\nu}$ and
the general affine connection $\Gamma^\lambda_{\mu\nu}$ are secondary, defined
through the first pair.

The quantity $D_\mu^{AB}e^B_\nu\equiv(D_\mu e_\nu)^A$ is a vector in the flat space, therefore
it can be decomposed in the frame field $e^A$ which forms a basis in the flat space.
We denote the expansion coefficients by $\Gamma^\lambda_{\mu\nu}$,
\beq
(D_\mu e_\nu)^A\; \bydef \;\Gamma^\lambda_{\mu\nu}\,e^A_\lambda\,,
\la{Gamma-def}\eeq
which serves as a definition of the affine connection $\Gamma^\lambda_{\mu\nu}$. It is equal to the sum
of the standard Christoffel symbol (or Levi-Civita connection) to which the general affine connection
reduces in the zero-torsion limit,
\beq
\bar \Gamma^\lambda_{\mu\nu} = \frac{1}{2}g^{\lambda\kappa}\left(\partial_\mu g_{\kappa\nu}
+\partial_\nu g_{\kappa\mu}-\partial_\kappa g_{\mu\nu}\right),
\la{Christoffel}\eeq
and the torsion part,
\begin{eqnarray}
\Gamma^\lambda_{\mu\nu}=\bar\Gamma^\lambda_{\mu\nu}
+T_{~\mu\nu}^{\lambda}+T_{~\nu\mu}^{\lambda}+T_{\mu\nu}^{~~\lambda}.
\la{Gamma-full}\end{eqnarray}

\subsection{Affine curvature and Riemann tensor}

\Eq{Gamma-def} can be rewritten as
\beq
\left(\nabla_\mu\right)^\kappa_\lambda\,e^{A\lambda}=-\omega^{AB}_\mu\,e^{B\kappa},
\la{covderR1}\eeq
where
\beq
\left(\nabla_\mu\right)^\kappa_\lambda\; \bydef \;\partial_\mu\,\delta^\kappa_\lambda+\Gamma^\kappa_{\mu\lambda}
\la{covderR2}\eeq
is the standard affine covariant derivative. The commutator of two covariant derivatives defines the affine
curvature tensor
\beq
\left[\nabla_\mu\nabla_\nu\right]^\kappa_\lambda\; \bydef \; R^\kappa_{~\lambda,\mu\nu}\;=\;
\partial_\mu\Gamma^\kappa_{\nu\lambda}-\partial_\nu\Gamma^\kappa_{\mu\lambda}
+\Gamma^\kappa_{\mu\rho}\Gamma^\rho_{\nu\lambda}-\Gamma^\kappa_{\nu\rho}\Gamma^\rho_{\mu\lambda},
\la{Riemann-curv1}\eeq
which is related to the curvature \ur{curvAB} built from the spin connection:
\beq
R^\kappa_{~\lambda,\mu\nu}=e^{A\kappa}\,e^B_\lambda\,{\cal F}^{AB}_{\mu\nu}\,.
\la{rel2}\eeq
The all-indices-down curvature tensor and the generalization of the Ricci tensor are
\beq
R_{\kappa\lambda,\mu\nu}=e^A_\kappa\,e^B_\lambda\,{\cal F}^{AB}_{\mu\nu},\qquad
R_{\kappa\lambda}\;\bydef \;R_{\kappa\mu,\lambda\nu}\,g^{\mu\nu}=e^A_\kappa\,e^{B\nu}\,{\cal F}^{AB}_{\lambda\nu}\,.
\la{rel3}\eeq

Let us denote with the bar the curvature defined by \Eq{Riemann-curv1} but built from the
symmetric Christoffel symbol $\bar\Gamma$ \ur{Christoffel}. $\bar R_{\kappa\lambda,\mu\nu}$
is then the standard, zero-torsion Riemann tensor satisfying the following relations:
\bea\n
&&\bar R_{\kappa\lambda,\mu\nu}=\bar R_{\mu\nu,\kappa\lambda}\,,\\
\n
&&\bar R_{\kappa\lambda,\mu\nu}+\bar R_{\kappa\mu,\nu\lambda}+\bar R_{\kappa\nu,\lambda\mu}=0\,,\\
&&\bar R_{\kappa\lambda,\mu\nu}\,\eps=0,\qquad \bar R_{\kappa\lambda}=\bar R_{\lambda\kappa}.
\la{rel-bar}\eea
These relations are not valid in the general case for the un-barred curvature $R_{\kappa\lambda,\mu\nu}$
\ur{rel3} if torsion is nonzero.

\section{The fermionic action}

In this section, we construct all possible bilinear fermion actions with zero and one covariant derivatives.
To make sure that we do not miss any terms, we prefer to use the two-component formalism, see Section III.A.
Any action is, in principle, allowed that is {\it i}) diffeomorphism-invariant and {\it ii}) invariant under
local Lorentz transformations \ur{Weyl_transf}. The first requirement means that, if only covariant (lower)
indices are used for $e^\pm_\mu, \nabla^{\pm\mp}_\mu, F^{\pm\mp}_{\mu\nu}$, they must be all contracted with
the antisymmetric $\eps$, in order to compensate for the change of coordinates in the volume element $d^4x$.
The second requirement means that, because of the gauge transformation laws \urss{Weyl_transf}{transfe}{transfnabla},
one has to alternate subscripts `plus' and `minus' in the chain. Fermion operators are at the ends of the
chain, such that the `plus' is followed by the Weyl field $\psi$ whereas the `minus' is followed by the
Weyl field $\chi$. On the contrary, $\psi^\dagger$ is always followed by a `minus' whereas $\chi^\dagger$
is followed by a `plus'.

\subsection{Zero-derivative terms}

The zero-derivative fermion action can have only two structures:
\begin{eqnarray}
S^f_0&=&\frac{1}{4!}\int\! d^4x\, i\,\eps\left[f_{01}(\psi^\dagger e^-_\kappa e^+_\lambda e^-_\mu e^+_\nu \psi)
+f_{02} (\chi^\dagger e^+_\kappa e^-_\lambda e^+_\mu e^-_\nu \chi)\right]\\
\nn
&=& \int\!d^4 x\,i\,\det(e)\left[f_{01}(\psi^\dagger\psi)-f_{02}(\chi^\dagger\chi)\right].
\end{eqnarray}
Both terms are Hermitian if $f_{01}$ and $f_{02}$ are real, since $\psi,\psi^\dagger$ and
$\chi,\chi^\dagger$ are Grassmann variables. The two terms transform into one another under parity
transformation. Therefore, if parity is not broken $f_{01}=-f_{02}=m$. In this case the two
terms combine into the mass term:
\beq
S^f_0=i\int\!d^4 x\,\det(e)\,m\,\Psi^\dagger \Psi.
\la{mass-term}\eeq

\subsection{One-derivative terms}

Since all covariant indices must be contracted with $\eps$, the total number of covariant indices
belonging to $e^\pm_\mu, \nabla^{\pm\mp}_\mu$ must be four. In one-derivative terms, one covariant
index belongs to $\nabla^{\pm\mp}_\mu$, therefore the number of frame fields must be three.
It means, first of all, that the number of `pluses' and `minuses' is odd, therefore all possible
one-derivative terms are necessarily off-diagonal in the Weyl fields $\psi,\chi$.
It is sufficient to consider operators of the type $(\chi^\dagger\ldots\psi)$ since the opposite order
$(\psi^\dagger\ldots\chi)$ will be obtained by hermitian conjugation.

{\it A priori} one can construct many terms satisfying these constraints, however all of them can be
reduced, using the algebra from Section III.A, to the following four terms:
\bea\n
&&\epsilon^{\kappa\lambda\mu\nu}(\chi^\dagger e^+_\kappa e^-_\lambda e^+_\mu \nabla^{-+}_\nu \psi),~~~
\epsilon^{\kappa\lambda\mu\nu}(\chi^\dagger e^+_\kappa e^-_\lambda \nabla^{+-}_\mu e^+_\nu \psi),\\
&&\epsilon^{\kappa\lambda\mu\nu}(\chi^\dagger e^+_\kappa \nabla^{-+}_\lambda e^-_\mu e^+_\nu \psi),~~~
\epsilon^{\kappa\lambda\mu\nu}(\chi^\dagger \nabla^{+-}_\kappa e^+_\lambda e^-_\mu e^+_\nu \psi).
\la{four-terms}\eea

The covariant derivative in these strings acts either on the Weyl field or on the tetrad.
One can commute the derivative to the utmost right position, therefore the only term with the
derivative of the fermion field is the usual Dirac term,
$\epsilon^{\kappa\lambda\mu\nu}(\chi^\dagger e^+_\kappa e^-_\lambda e^+_\mu \nabla^{-+}_\nu \psi)$.
All other terms contain derivatives of the tetrad, which should be antisymmetrized in the covariant
indices to make world tensors under diffeomorphisms. This forms the torsion tensor \ur{torsion}.
Therefore, apart from the Dirac term, \Eq{four-terms} describes the following three structures:
\beq
\eps\,(\chi^\dagger T_{\mu\nu}^+ e^-_\kappa e^+_\lambda \psi),\qquad
\eps (\chi^\dagger e^+_\kappa T_{\mu\nu}^- e^+_\lambda \psi),\qquad
\eps (\chi^\dagger e^+_\kappa e^-_\lambda T_{\mu\nu}^+ \psi)\,
\la{three-terms}\eeq
where $T_{\mu\nu}^\pm=T^A_{\mu\nu}\sigma_A^\pm$. In addition to the four terms in \Eq{four-terms} one can
consider
$$
\eps(\chi^\dagger e^+_\mu\psi)\, \Tr(T^+_{\kappa\lambda} e^-_\nu)\quad {\rm and}\quad
\eps\chi^\dagger e^+_\mu\psi\, \Tr(T^-_{\kappa\lambda} e^+_\nu).
$$
These two terms are in fact identical and equal to
$\half\eps\((\chi^\dagger e^+_\kappa e^-_\lambda T^+_{\mu\nu}\psi)-(\chi^\dagger T^+_{\mu\nu}e^+_\kappa e^-_\lambda \psi)\)$
belonging to the set \ur{three-terms}. Further on, the three terms in \Eq{three-terms} are not independent as there is
an algebraic identity
$$
\epsilon^{\kappa\lambda\mu\nu}\(T_{\mu\nu}^+ e^-_\kappa e^+_\lambda
+2 e^+_\kappa T_{\mu\nu}^- e^+_\lambda + e^+_\kappa e^-_\lambda T_{\mu\nu}^+\) \equiv 0
$$
leaving us with only two terms with the torsion field, say, the first and the last term in \ur{three-terms}.

In principle, one can build diffeomorphism-invariant actions using {\em two} Levi-Civita symbols $\eps$ contracted
with one $\nabla^{\pm\mp}_\mu$ and seven $e^\pm_\mu$ but {\em divided}  by $\det(e)$ such that the expression
is again invariant under the change of coordinates, together with the volume element. However, all such
expressions are in fact identical to linear combinations of invariants listed in \ur{four-terms}.

The resulting three independent fermion actions can be presented in a more simple form. We
use the decomposition of the torsion tensor \ur{T-split} and notice that actually only 8 out of the possible 24
components of the torsion field couple to fermions in this order: the traceless symmetric part $t_{\mu\nu}^{~~\lambda}$
decouples. Indeed, one has:
\begin{eqnarray*}
&&\eps(\chi^\dagger e^+_\kappa e^-_\lambda e^+_\mu \nabla^{-+}_\nu \psi)
=-6 \det(e)(\chi^\dagger e^{+\mu} \nabla_\mu^{-+} \psi),\\
&&\epsilon^{\kappa\lambda\mu\nu}(\chi^\dagger T_{\mu\nu}^+ e^-_\kappa e^+_\lambda \psi)
=-8\det(e)\,\left(-\frac{1}{2}v_\mu+a_\mu\right)(\chi^\dagger e^{+\mu} \psi), \\
&&\epsilon^{\kappa\lambda\mu\nu}(\chi^\dagger e^+_\kappa e^-_\lambda T_{\mu\nu}^+ \psi)
=-8\det(e)\,\left(-\frac{1}{2}v_\mu-a_\mu\right)(\chi^\dagger e^{+\mu} \psi).
\end{eqnarray*}
Therefore, the most general one-derivative fermion action has the form
\beq
S^f_1=\int\!d^4 x\,\det(e)\left[f_{10} (\chi^\dagger e^{+\lambda}\nabla_\lambda^{-+}\psi)
+f_{11}\,a_\lambda(\chi^+ e^{+\lambda} \psi) +f_{12}\, v_{\lambda}(\chi^\dagger e^{+\lambda}\psi)\right]+{\rm h.c.},
\la{Sf11}\eeq
where $f_{1i}$ are arbitrary complex numbers.

It should be noted that the operator $\nabla_\lambda^{-+}$ in the first term contains the full
spin connection $\omega_\mu$ which, according to \Eq{omega-full}, can be written as a sum of the
zero-torsion part $\bar\omega_\mu$ \ur{omega-bar} and terms proportional to torsion.
The difference of two operators is from \urs{omega-full}{T-split} identically
\beq
e^{\pm\,\lambda}\left(\nabla^{\mp\pm}_\lambda - \bar{\nabla}^{\mp\pm}_\lambda\right)
=e^{\pm\lambda}(-v_\lambda\mp a_\lambda)
\la{rel1}\eeq
where $\bar{\nabla}^{\mp\pm}_\lambda$ is the covariant derivative computed in the zero-torsion limit.
We have
\beq
S^f_1=\int\!d^4 x\,\det(e)\left[g_0 (\chi^\dagger e^{+\lambda}\bar\nabla_\lambda^{-+}\psi)
+g_1\,a_\lambda(\chi^\dagger e^{+\lambda} \psi) +g_2\, v_{\lambda}(\chi^\dagger e^{+\lambda}\psi)\right]+{\rm h.c.}\,,
\la{Sf11bar}\eeq
where the new couplings are
$g_0=f_{10},\quad g_1=f_{11}-f_{10},\quad g_2=f_{12}-f_{10}.
$

\subsection{Hermitian action}

We now add explicitly the Hermitian conjugate action in \Eq{Sf11bar}. It is straightforward for
the second and third terms; changing the order of the fermion operators brings the minus sign.
The Hermitian conjugation of the first term is more involved. We first take its complex conjugate,
interchange the order of fermion operators $(\chi,\psi^*)$, integrate by parts and get
$$
\det(e)\(\chi^\dagger e^{+\lambda}\bar\nabla_\lambda^{-+}\psi\)_{\rm h.c.}
=\(\psi^\dagger\bar\nabla_\lambda^{-+}e^{-\lambda}\det(e)\,\chi\).
$$
We next write $e^{-\lambda}\det(e)=-(1/6)\epsilon^{\lambda\alpha\beta\gamma}\,e^-_\alpha e^+_\beta e^-_\gamma$
and drag $\bar\nabla_\lambda^{-+}$ to the right through this expression. Since torsion in $\bar\nabla_\lambda^{-+}$
is by construction zero, it commutes with the tetrads owing to \Eq{torsion}, and we get
$$
\(\psi^\dagger\bar\nabla_\lambda^{-+}e^{-\lambda}\det(e)\,\chi\)
=\(\psi^\dagger e^{-\lambda}\det(e)\,\bar\nabla_\lambda^{+-}\chi\).
$$
We, thus, obtain an explicitly Hermitian fermion action written in the 2-component spinor Weyl form
\begin{eqnarray}
S^f_1 = \int\!d^4x \!\!\!&&\det(e) \left[g_0(\chi^\dagger e^{+\mu}\bar\nabla_\mu^{-+}\psi)
+g_0^*(\psi^\dagger e^{-\mu}\bar\nabla_\mu^{+-}\chi)\right.\\
\nn
&&+\left.a_\mu\left(g_1(\chi^\dagger e^{+\mu}\psi)-g_1^*(\psi^\dagger e^{-\mu}\chi)\right)
+v_\mu\left(g_2(\chi^\dagger e^{+\mu}\psi)-g_2^*(\psi^\dagger e^{-\mu}\chi)\right)\right].
\label{S-short}\end{eqnarray}

The constant $g_0$ can be made real by redefining the overall phases of $\psi,\psi^\dagger$ and $\chi,\chi^\dagger$.
Indeed, if the argument of $g_0$ is $\alpha$ ($g_0=|g_0|e^{i\alpha}$), the phase rotation
$\psi\to\exp(-i\alpha/2)\psi,\;\psi^\dagger\to\psi^\dagger\exp(i\alpha/2),\;
\chi\to\exp(i\alpha/2)\chi,\;\chi^\dagger\to\chi^\dagger\exp(-i\alpha/2)$ obviously makes $g_0$ real,
and it can be further on put to unity by rescaling of the $\psi,\chi$ fields.
Therefore, we can put $g_0=1$ to make the Dirac kinetic energy term standard.

Finally, recalling the definition of the bi-spinors \ur{bi-spinor} and the Dirac matrices \ur{Dirac-gamma}
we rewrite the action \ur{S-short} in the 4-component Dirac form:
\beq
S^f_1=  \int\! d^4 x\,\det(e)\,\Psi^\dagger \gamma^\mu \left[\bar {\cal D}_\mu
+ a_\mu(g_1^-+g_1^+\gamma_5)+v_\mu(g_2^-+g_2^+\gamma_5)\right]\Psi
\la{S-gamma}\eeq
where
\beq
g_{1,2}^\pm=\frac{g_{1,2}\pm g^*_{1,2}}{2},\qquad \gamma^\mu=e^{A\mu} \gamma_A,\qquad
\bar {\cal D}_\mu =\partial_\mu {\bf 1}+\frac{1}{8}[\gamma_A\gamma_B]\bar{\omega}_\mu^{AB}.
\la{g}\eeq
The action \ur{S-gamma} is by construction and manifestly Hermitian.

In the `minimal model' often discussed in the literature~\cite{Hehl:1976kj,Hammond:2002rm,Perez:2005pm},
the only source of the fermion interaction with torsion is the Dirac term \ur{ferm-gen} with the full
spin connection including its torsion part. In this case $g^+_1=-1$, and all other constants are zero,
therefore only the axial part of the torsion couples to fermions. The term proportional to $g_2^-$
was first considered in Ref.~\cite{Freidel:2005sn} although in another form, see also its discussion
in Ref.~\cite{Khriplovich:2005jh}. These authors take the following Lagrangian generalizing the `minimal model':
$$
S^f= \int\! d^4 x\,\det(e)\(\frac{1-i\alpha}{2}\,\Psi^\dagger \gamma^\mu {\cal D}_\mu\Psi
-\frac{1+i\alpha}{2}\,({\cal D}_\mu\Psi)^\dagger \gamma^\mu \Psi\).
$$
Integrating the second term by parts we bring this Lagrangian to our form \ur{S-gamma} with the particular values of the constants: $g_1^+=-1$, $g_2^-=\alpha$, the rest being zero.

The complete list of four one-derivative fermion-torsion actions \ur{S-gamma} was presented in Ref.~\cite{Kostelecky:2007kx} where in addition nine terms with one extra derivative were suggested.

\subsection{From Euclidean to Minkowski signature}

The standard dictionary translating Euclidean into Minkowski variables (see, {\it e.g.}~\cite{Vainshtein:1981wh})
reads:
$$
%x^0=-ix^4_E,~~~x^i=x^i_E,~~~e^0_0=-e^4_4,~~~e^0_i=ie^4_{iE},~~~e^a_0=ie^a_{4E},~~~ e^a_i=e^a_{iE},
\gamma^0=\gamma_{4E},~~~\gamma^i=-i\gamma^i_E,~~~\Psi=\Psi_E,~~~\bar\Psi=i\Psi^\dagger_E,~~~\gamma^5=i\gamma_{5E},~~~S=i S_E.
$$
Therefore, we obtain from \Eqs{mass-term}{S-gamma} the full Minkowski fermion action with zero and one derivative:
\beq
S^f = \int\! d^4 x\,\det(e)\,\bar\Psi\left[i\gamma^\mu \left(\bar {\cal D}_\mu
+ a_\mu(g_1^- -ig_1^+\gamma^5)+v_\mu(g_2^- -ig_2^+\gamma^5)\right)-m\right]\Psi\,.
\la{S-gamma-M}\eeq
Assuming $a_\mu$ is an axial and $v_\mu$ is a vector field, the terms with $g_1^-$ and $g_2^+$ break
P-parity; other terms are parity-even. We remind that for a quantum amplitude one takes $\exp(iS)$.

\section{Torsion as an Abelian gauge field}

If parity is conserved, one has to put $g_1^- = g_2^+ =0$ in \Eq{S-gamma-M}. We denote the two nonzero constants
that are left as $g_1^+ =g_a$ (it is real by construction \ur{g}) and $g_2^- = -i g_v$ (it is purely imaginary),
and rewrite the fermion Lagrangian with torsion as
\beq
\bar\Psi\gamma^\mu\left(\bar D_\mu-il_\mu\,\frac{1+\gamma^5}{2}-ir_\mu\, \frac{1-\gamma^5}{2}\right)\Psi,\qquad
l_\mu = g_v\,v_\mu +g_a\,a_\mu,\qquad  r_\mu = g_v\,v_\mu - g_a\,a_\mu.
\la{lr}\eeq
This Lagrangian is clearly invariant under the Abelian $U(1)_{\rm L}\times U(1)_{\rm R}$ gauge transformation
\beq
\Psi_{\rm L}=\psi \to e^{i\alpha(x)}\,\psi,\qquad \Psi_{\rm R}=\chi \to e^{i\beta(x)}\,\chi,\qquad
l_\mu \to l_\mu+\partial_\mu\alpha, \qquad r_\mu \to r_\mu+\partial_\mu\beta.
\la{U1-gauge}\eeq
This invariance, in the $U(1)_{\rm V}\times U(1)_{\rm A}$ form, has been previously noticed in
Ref.~\cite{deBerredoPeixoto:1999vj}.

Therefore, the inclusion of torsion is equivalent to promoting the fermion part of the standard gravity
invariant under the Lorentz $SU(2)_{\rm L}\times SU(2)_{\rm R}$ gauge group to being invariant under
the larger $U(2)_{\rm L}\times U(2)_{\rm R}$ gauge group!

If this gauge symmetry is preserved by the bosonic part of the action, it has to depend only on
the curvatures $f^{({\rm L})}_{\mu\nu}=\partial_\mu l_\nu-\partial_\nu l_\mu,\;
f^{({\rm R})}_{\mu\nu}=\partial_\mu r_\nu-\partial_\nu r_\mu$. Linear terms are zero, so
the expansion starts with quadratic terms in $f_{\mu\nu}$. The vector part, $v_\mu$, is then identical
to the photon. It should be mentioned that, since it is an Abelian field, one is free to ascribe arbitrary
coupling constants or `charges' with which this field interacts with various fermion species.
To be separated from the photon, this field has to have other `charges' with respect to
fermion species, and be massive. The only thing we know is that its mass must be larger than the
experimental restriction on neutral intermediate bosons, of the order of $1\,{\rm TeV}$. The same
applies to the $U(1)$ axial boson $a_\mu$ or their linear combinations $l_\mu,\; r_\mu$.

The appearance of mass terms for $l_\mu,r_\mu$ fields means breaking of the $U(1)_{\rm L}\times U(1)_{\rm R}$
gauge symmetry, which can be either explicit or spontaneous by some kind of a Higgs effect. As we shall see
in the next section, adding terms quadratic in torsion implies explicit breaking of the
$U(1)_{\rm L}\times U(1)_{\rm R}$ symmetry. In principle, there is nothing wrong about it as the gauge
symmetry is Abelian. Probably, spontaneous breaking would be more aesthetic but in the absence of the
microscopic theory we can only speculate about it.

From the viewpoint that torsion fields $a_\mu,v_\mu$ are just another set of gauge vector bosons interacting
with fermions, we do not see compelling reasons why their masses should be of the order of the Planck mass,
as suggested by the `gravitational' approach to torsion: with our present lack of deeper understanding the
masses can be anything beyond the phenomenologically established limits~\cite{Belyaev:2007fn}.

If parity is not conserved (meaning $g_1^-,g_2^+$ are nonzero), one can still consider $l_\mu$ and $r_\mu$
as compensating gauge fields. However, then they have to be complex, and compensate point-dependent real
dilatations of the chiral fields $\psi,\chi$ and not only their phases. A discussion of this interesting
topic lies beyond the scope of the paper.

\section{The bosonic action}

Taking a purely phenomenological stand, one may inquire what terms in the bosonic action can be
written that preserve {\it i}) diffeomorphism-invariance and {\it ii}) invariance under the gauge
Lorentz group. In this section, we give the full list of invariants quadratic in torsion, invariants
quadratic in curvature, and invariants that are linear in torsion but containing $\bar\nabla \bar R$.

\subsection{Invariants quadratic in torsion}

A general way to construct quadratic invariants is to consider the following invariant under the diffeomorphism,
\beq
K^{A[CD],B[EF]} = \det(e)\,(T^A_{\mu\nu}\,e^{C\mu}\,e^{D\nu})(T^B_{\alpha\beta}\,e^{E\alpha}\,e^{F\beta}),
\la{invK}\eeq
and to contract the flat indices into a Lorentz-group scalar. Since torsion is antisymmetric in world indices,
this expression is antisymmetric in $[CD]$ and $[EF]$, meaning that the pairs of the frame fields belong
to the ${\bf 6}$-dimensional representation of the $SO(4)$ Lorentz group or to the $(\bf{3,1}) \oplus ({\bf 1,3})$
representation of the $SU(2)_{\rm L}\times SU(2)_{\rm R}$ group, while $T^A$ belongs to the
$(\bf{2,2})$ representation of that group.

The direct product of $T^A$ and $e^{[C}e^{D]}$ belongs to the $({\bf 2,2})\otimes \left((\bf{3,1}) \oplus ({\bf 1,3})\right)
= ({\bf 2,2})_1\oplus ({\bf 2,2})_2\oplus ({\bf 2,4})\oplus ({\bf 4,2})$ representations, which should be multiplied
by the same combination. There are 5 singlets arising from
$$
({\bf 2,2})_1\otimes ({\bf 2,2})_1,\quad ({\bf 2,2})_1\otimes ({\bf 2,2})_2,\quad ({\bf 2,2})_2\otimes ({\bf 2,2})_2,\quad
({\bf 2,4})\otimes ({\bf 2,4}),\quad ({\bf 4,2})\otimes ({\bf 4,2}).
$$
Therefore, there are precisely five linear independent invariants which we write as
\bea\n
K_1 &=& \det(e)\;T^A_{\mu\nu}\;T^A_{\alpha\beta}\;g^{\mu \alpha}\;g^{\nu \beta}
~~\,~~~~=\;\det(e)\(-\frac{8}{3}\,a^\mu a_\mu+\frac{2}{3}\,v^\mu v_\mu +\frac{4}{9}\,t_{\mu\nu}^{~~\lambda}\,t_{~~\lambda}^{\mu\nu}\) , \\
\n
K_2 &=& \det(e)\;T^A_{\mu\nu}\;T^B_{\alpha\beta}\;e^{A\mu}\;e^{B\alpha}\;g^{\nu \beta}
~=\;\det(e)\,v^\mu v_\mu, \\
\n
K_3 &=& \det(e)\;T^A_{\mu\nu}\; T^B_{\alpha\beta}\; e^{A\alpha} e^{B\mu}\; g^{\nu \beta}
~\;=\;\det(e)\(\frac{8}{3}\,a^\mu a_\mu+\frac{1}{3}\,v^\mu v_\mu +\frac{2}{9}\,t_{\mu\nu}^{~~\lambda}\,t_{~~\lambda}^{\mu\nu}\), \\
\n
K_4 &=& \frac{1}{2} \epsilon^{\mu\nu\alpha\beta}\;T^A_{\mu\nu}\;T^A_{\alpha\beta}
~~~~~~~~~~~~~~~\,\,=\;\det(e)\,\frac{8}{3}\,a^\mu v_\mu
+\frac{2}{9}\,t_{\alpha\beta}^{~~\lambda}\,t_{\mu\nu,\lambda}\;\epsilon^{\alpha\beta\mu\nu}, \\
K_5 &=& \epsilon^{\mu\alpha\kappa\lambda}\;T^A_{\mu\nu}\;T^B_{\alpha\beta}\;e^A_\kappa\;e^B_\lambda\;g^{\nu\beta}
%= \det(e)\;\epsilon^{ABCD}\;T^A_{\mu\nu}\;T^B_{\alpha\beta}\;e^{C\mu}\;e^{D\alpha}\;g^{\nu \beta}.
~~~~~=\;\det(e)\,\frac{8}{3}\,a^\mu v_\mu-\frac{1}{9}\,t_{\alpha\beta}^{~~\lambda}\,t_{\mu\nu,\lambda}\;\epsilon^{\alpha\beta\mu\nu}.
\la{5-invariants}\eea

In the last column we used the decomposition of the torsion tensor \ur{T-split}.
The last two terms are $P,T$-odd; the first three are even. The first four terms have been known for a
long time as they emerge from the leading Einstein--Cartan terms \ur{bos-lowest}, see below.
To the best of our knowledge, the fifth invariant appears for the first time in the very recent Ref.~\cite{Baekler:2010fr}.

We now recall that in the Einstein--Cartan formulation there are two leading terms linear in the curvature,
see the second and the third terms in \Eq{bos-lowest}. Following the general strategy we split them into a piece
that survives in the zero-torsion limit, plus corrections from torsion. One has~\cite{Sezgin:1979zf,Nieh:1981ww,Mielke:2009zz}:
\bea\la{EC1}
\frac{1}{4}\,\epsilon^{\mu\nu\alpha\beta}\,\epsilon^{ABCD}\,{\cal F}^{AB}_{\mu\nu}\,e_{\alpha}^C\,e_{\beta}^D
&=&\sqrt{-g}\,\bar R +K_1 - 4K_2 + 2\,K_3
+ 4\,\partial_{\mu}\left(\sqrt{-g}\,v^\mu \right),\\
\la{EC2}
\frac{1}{2}\,\eps\,{\cal F}^{AB}_{\mu\nu}\;e_{\kappa}^A\;e_{\lambda}^B
&=&2K_4-4\partial_{\mu} \left(\sqrt{-g}\,a^\mu \right).
\eea

We see thus that four out of possible five terms quadratic in torsion are induced by terms linear in curvature,
with concrete coefficients. We shall, however, consider the general case where the terms $K_{1...5}$ are included
in the bosonic part of the action with arbitrary real coefficients $k_{1...5}$:
\bea\la{bos-2}
S^b_2\!&=&\int\!d^4x\;\frac{1}{2}\sum_{m=1}^5 k_m\,K_m(T) \\
\n
&=\!&\!\!\int\!\!d^4x \det(e)\frac{1}{2}\left(\!M^2_{aa}\,a^\mu a_\mu +2M^2_{av}\,a^\mu v_\mu+M^2_{vv}\,v^\mu v_\mu
+M^2_{tt}\,t_{\mu\nu}^{~~\lambda} t^{\mu\nu}_{~~\lambda}
+\frac{1}{2}\,M^2_{\epsilon tt}\;t_{\alpha\beta}^{~~\lambda}\,t_{\mu\nu,\lambda}\,
\frac{\epsilon^{\alpha\beta\mu\nu}}{\det(e)}\!\right).
\eea
The new constants having the meaning of the masses squared of the torsion fields are linear combinations of the original constants:
$$
M^2_{aa}=\frac{8(k_3-k_1)}{3},~~~M^2_{av}=\frac{4(k_4+k_5)}{3},
~~~M^2_{vv}=\frac{2k_1+3k_2+k_3}{3},
$$
$$
M^2_{tt}=\frac{2(2k_1+k_3)}{9},~~~~M^2_{\epsilon tt}=\frac{2(2k_4-k_5)}{9}.
$$
The second and fifth terms are P,T-odd, the rest are even. The first three terms in \Eq{bos-2} are mass terms for the $a_\mu,v_\mu$ bosons or for their linear combinations
$l_\mu,r_\mu$, that break explicitly the $U(1)\times U(1)$ gauge symmetry discussed in Section V.

The system must be stable with respect to small low-momenta fluctuations of torsion about the flat space. It means that all eigenvalues of the mass matrix \ur{bos-2} must be
positive. This condition requires that
\beq
M^2_{vv}+M^2_{aa}>0,\qquad M^2_{vv}M^2_{aa}>M^4_{av},\qquad M^2_{tt}>M^2_{\epsilon tt}
\la{stability}\eeq
which is satisfied in a broad range of the constants $k_{1\!-\!5}$. See also the discussion of the positivity of the mass matrix in Ref.~\cite{Baekler:2010fr}.

In the `minimal model' corresponding to extracting torsion terms from the leading-order action \ur{bos-lowest} only, see \Eqs{EC1}{EC2}, one obtains
\bea\n
&& M^{2\;{\rm min}}_{aa}=-\frac{8}{3}\frac{M_P^2}{16\pi},\qquad M^{2\;{\rm min}}_{vv}=\frac{8}{3}\frac{M_P^2}{16\pi},
\qquad M^{2\;{\rm min}}_{tt}=-\frac{8}{9}\frac{M_P^2}{16\pi},\\
&& M^{2\;{\rm min}}_{av}=-\iota\frac{8}{3}\,\frac{M_P^2}{16\pi},\qquad
M^{2\;{\rm min}}_{\epsilon tt}=-\iota\,\frac{8}{9}\,\frac{M_P^2}{16\pi},
\la{MMmin}\eea
where the {\em iota} parameter $\iota$ is the coefficient in front of the $P,T$-odd action \ur{EC2}. We coincide in this table of masses with Ref.~\cite{Mercuri:2006um}, after adjusting the normalization.

The eigenvalues of mass-squared matrix for $a_\mu,v_\mu$ are $\pm\sqrt{1+\iota^2}(8/3)M_P^2/16\pi$.
A check of the above algebra is that at purely imaginary values $\iota=\pm i$ the eigenvalues are zero. Indeed, at these values the self-dual or anti-self-dual combination
${\cal F}^{AB}\pm i \half \epsilon^{ABCD}\,{\cal F}_{CD}$ drops out of the action \ur{bos-lowest}.

At real values of the {\it iota} parameter one of the eigenvalues is always negative. It means that the path integral over $a_\mu,v_\mu$ fields strictly speaking does not exist, therefore the `minimal model' cannot be complete.

\subsection{Invariants quadratic in curvature}

Such terms arise from the diffeomorphism-invariant structure
\beq
G^{[AB][CD][EF][GH]}=\det(e)\,{\cal F}^{AB}_{\alpha\beta}\,{\cal F}^{CD}_{\gamma\delta}\,
e^{E\alpha}\,e^{F\beta}\,e^{G\gamma}\,e^{H\delta}
\la{G0}\eeq
belonging to the ${\bf 6}\otimes{\bf 6}\otimes{\bf 6}\otimes{\bf 6}$ representation of the Lorentz
group, out of which one can extract 10 independent Lorentz-group invariants. Here is their list,
expressed through the full Riemann tensor \ur{rel3}:
\bea\la{G}
G_1 &=& \frac{1}{16\det(e)}\left( \epsilon^{\mu\nu\alpha\beta}\,\epsilon^{ABCD}\,
{\cal F}^{AB}_{\mu\nu}\,e_{\alpha}^C\,e_{\beta}^D \right)^2  = \sqrt{-g}\,R^2 ,\\
\n
G_2 &=& \frac{1}{4\det(e)}\,{\cal F}^{AB}_{\mu\nu} \, {\cal F}^{AB}_{\alpha\beta}\,\epsilon^{\mu\nu\lambda\rho}\,
\epsilon^{\alpha\beta\gamma\delta}\,g_{\lambda\gamma}\,g_{\rho\delta}
=\sqrt{-g}\,R_{\lambda\rho,\mu\nu}\,R^{\lambda\rho,\mu\nu} ,\\
\n
G_3 &=& \frac{1}{16\det(e)}\,{\cal F}^{AB}_{\mu\nu} \, {\cal F}^{CD}_{\alpha\beta} \,
\epsilon^{\mu\nu\lambda\rho}\,\epsilon^{\alpha\beta\gamma\delta}\, \epsilon^{ABEF}\,
\epsilon^{CDGH}\,e_{\gamma}^E\,e_{\delta}^F\,e_{\lambda}^G\,e_{\rho}^H
= \sqrt{-g}\,R_{\lambda\rho,\mu\nu}\,R^{\mu\nu,\lambda\rho} ,\\
\n
G_4 &=& \frac{1}{4}\,\epsilon^{\mu\nu\alpha\beta}\,\epsilon^{ABCD}\,
{\cal F}^{AB}_{\mu\nu}\,{\cal F}^{CD}_{\alpha\beta}
= \sqrt{-g} \left( R^2 - 4\,R_{\lambda\mu}\,R^{\mu\lambda}
+ R_{\lambda\rho,\mu\nu}\,R^{\mu\nu,\lambda\rho} \right) ,\\
\n
G_5 &=& \frac{1}{4\det(e)}\,{\cal F}^{AB}_{\mu\nu} \, {\cal F}^{CD}_{\alpha\beta}\,\epsilon^{\mu\nu\lambda\rho}\,
\epsilon^{\alpha\beta\gamma\delta}\,e_{\gamma}^A\,e_{\delta}^B\,e_{\lambda}^C\,e_{\rho}^D
= \sqrt{-g} \left( R^2 - 4\,R_{\lambda\mu}\,R^{\lambda\mu}+ R_{\lambda\rho,\mu\nu}\,R^{\lambda\rho,\mu\nu} \right) ,\\
\n
G_6 &=& \frac{1}{\det(e)} \left( \epsilon^{\mu\nu\alpha\beta}\,{\cal F}^{AB}_{\mu\nu}\,e_{\alpha}^A\,e_{\beta}^B \right)^2
= \frac{1}{\sqrt{-g}}\, \left( \epsilon^{\lambda\rho\mu\nu}\,R_{\lambda\rho,\mu\nu} \right)^2 ,\\
\n
G_7 &=& \frac{1}{4e}\,\left( \epsilon^{\mu\nu\alpha\beta}\,
\epsilon^{ABCD}\,{\cal F}^{AB}_{\mu\nu}\,e_{\alpha}^C\,e_{\beta}^D \right)\,
\left(\epsilon^{\mu\nu\alpha\beta}\,{\cal F}^{AB}_{\mu\nu}\,e_{\alpha}^A\,e_{\beta}^B \right)
= \epsilon^{\lambda\rho\mu\nu}\,R\,R_{\lambda\rho,\mu\nu} ,\\
\n
G_8 &=& \epsilon^{\mu\nu\alpha\beta}\,{\cal F}^{AB}_{\mu\nu}\,{\cal F}^{AB}_{\alpha\beta}
= \epsilon^{\mu\nu\alpha\beta}\,R_{\lambda\rho,\mu\nu}\,R^{\lambda\rho}_{~~\alpha\beta} ,\\
\n
G_9 &=& \frac{1}{4\det(e)}\,{\cal F}^{AB}_{\mu\nu}\,{\cal F}^{CD}_{\alpha\beta}\,\epsilon^{\mu\nu\lambda\rho}\,
\epsilon^{\alpha\beta\gamma\delta}\,\epsilon^{ABCD}\,g_{\lambda\gamma}\,g_{\rho\delta}
= \epsilon^{\lambda\rho\gamma\delta}\,R_{\lambda\rho,\mu\nu}\,R_{\gamma\delta}^{~~\mu\nu} ,\\
\n
G_{10} &=& \frac{1}{\det(e)}\,{\cal F}^{AB}_{\mu\nu}\,{\cal F}^{CD}_{\alpha\beta}\,\epsilon^{\mu\nu\lambda\rho}\,
\epsilon^{\alpha\beta\gamma\delta}\,\epsilon^{CDEF}\,e^A_{\gamma}\,e^B_{\delta}\,e^E_{\lambda}\,e^F_{\rho}
= \epsilon^{\lambda\rho\alpha\beta}\,R_{\lambda\rho,\mu\nu}\,R^{\mu\nu}_{~~\alpha\beta}\,.
\eea
Invariants 7-10 are $P,T$-odd, the rest are even. $G_4$ and $G_8$ are full derivatives even if torsion is non-zero. The $P,T$-even invariants $G_{1\!-\!6}$ have been first constructed by Neville~\cite{Neville:1978bk}.
%
%We note that in Ref.~\cite{Sezgin:1979zf} the invariants $G_{6,7,9,10}$ have been missed,
%and that may affect the conclusions made there.

In the zero-torsion limit one replaces 
$R_{\kappa\lambda,\mu\nu}\to \bar R_{\kappa\lambda,\mu\nu}$ which satisfies the relations \ur{rel-bar}. Therefore, in this limit one has $G_2=G_3$, $G_4=G_5$, $G_{6,7}=0$, $G_8=G_9=G_{10}$. Thus, in the zero-torsion limit one is left, apart from two full derivatives, with only two well-known invariants, namely
\beq
\sqrt{-g}\,\bar R^2\quad{\rm and}\quad \sqrt{-g}\,\bar R_{\kappa\lambda}\,\bar R^{\kappa\lambda}.
\la{RR}\eeq

\subsection{Invariants linear in torsion}

Using the covariant derivative of the curvature it is possible to construct invariants that are linear in torsion and linear in $\nabla R$. The general structure from which all invariants of this kind can be derived
is  
\bea
L^{AB[CD][EF][GH]} = \det(e)\,(\bar{\nabla}_\lambda \bar{R}_{\alpha\beta,\gamma\delta})\,T^A_{\mu\nu}\,
e^{B\lambda}\,e^{C\mu}\,e^{D\nu}\,e^{E\alpha}\,e^{F\beta}\,e^{G\gamma}\,e^{H\delta}
\la{L0}\eea
where $\bar\nabla$ is the covariant derivative with the no-torsion Christoffel symbol \ur{Christoffel}.
%$$
%\bar{\nabla}_\lambda \bar{R}_{\alpha\beta,\gamma\delta}
%= \partial_{\lambda} \bar{R}_{\alpha\beta,\gamma\delta}
%- \bar{\Gamma}^{\rho}_{\lambda\alpha}\;\bar{R}_{\rho\beta,\gamma\delta}
%- \bar{\Gamma}^{\rho}_{\lambda\beta}\;\bar{R}_{\alpha\rho,\gamma\delta}
%- \bar{\Gamma}^{\rho}_{\lambda\gamma}\;\bar{R}_{\alpha\beta,\rho\delta}
%- \bar{\Gamma}^{\rho}_{\lambda\delta}\;\bar{R}_{\alpha\beta,\gamma\rho}.
%$$
It belongs to the $ \bf{6}^3 \otimes \bf{4}^2 $ representation of the Lorentz group, which contains
20 singlets:
$$
\begin{array}{cccccc}
L'_{1} &=& \det(e)\;T^A_{\mu\nu}\;e_{\rho}^A\;\bar{\nabla}_\lambda \bar{R}^{\lambda\rho,\mu\nu}, &
L'_{2} &=& \det(e)\;T^A_{\mu\nu}\;e_{\rho}^A\;\bar{\nabla}_\lambda \bar{R}^{\mu\nu,\lambda\rho}, \\
L'_{3} &=& \det(e)\;T^A_{\mu\nu}\;e_{\rho}^A\;\bar{\nabla}^{[\mu} \bar{R}^{\nu]\rho}, &
L'_{4} &=& \det(e)\;T^A_{\mu\nu}\;e_{\rho}^A\;\bar{\nabla}^{[\mu} \bar{R}^{\rho\nu]}, \\
L'_5 &=& \det(e)\;T^A_{\mu\nu}\;e^{A\,[\mu}\;\bar{\nabla}^{\nu]} \bar{R}, &
L'_6 &=& \det(e)\;T^A_{\mu\nu}\;e_{\rho}^A\;\bar{\nabla}_\lambda \bar{R}^{\lambda[\mu,\nu]\rho}, \\
L'_7 &=& \epsilon^{\lambda\rho\mu\nu}\;T^A_{\mu\nu}\;e_{\rho}^A\;\bar{\nabla}_{\lambda} \bar{R}, &
L'_{8} &=& \epsilon^{\mu\nu\gamma\delta}\;
T^A_{\mu\nu}\;e_{\rho}^A\;\bar{\nabla}_{\lambda} \bar{R}^{\lambda\rho}_{~~\gamma\delta}, \\
L'_{9} &=& \epsilon^{\mu\nu\gamma\delta}\;T^A_{\mu\nu}\;e_{\rho}^A\;\bar{\nabla}_{\lambda} \bar{R}_{\gamma\delta}^{~~\lambda\rho}, &
L'_{10} &=& T^A_{\mu\nu}\;e^{A\,\alpha}\;g^{\gamma[\mu}\;
\epsilon^{\nu]\lambda\beta\delta}\;\bar{\nabla}_\lambda \bar{R}_{\alpha\beta,\gamma\delta}\,, \\
L'_{11} &=& T^A_{\mu\nu}\;e^{A\,\alpha}\;g^{\gamma[\mu}\;
\epsilon^{\nu]\lambda\beta\delta}\;\bar{\nabla}_\lambda \bar{R}_{\gamma\delta,\alpha\beta}\,, &
L'_{12} &=& \det(e)\;T^A_{\mu\nu}\;e^{A\,\lambda}\;\bar{\nabla}_{\lambda} \bar{R}^{\mu\nu}, \\
L'_{13} &=& \frac{1}{\det(e)} \epsilon^{\lambda\rho\alpha\beta}\epsilon^{\mu\nu\gamma\delta}
\;T^A_{\mu\nu}\,e^A_{\rho}\, \bar{\nabla}_\lambda \bar{R}_{\alpha\beta,\gamma\delta}, &
L'_{14} &=& \frac{1}{\det(e)} \epsilon^{\lambda\rho\mu\nu}\epsilon^{\alpha\beta\gamma\delta} \;
T^A_{\mu\nu}\,e^A_{\rho}\, \bar{\nabla}_\lambda \bar{R}_{\alpha\beta,\gamma\delta}, \\
%\end{array}
%$$
%
%$$
%\begin{array}{cccccc}
L'_{15} &=& \frac{1}{\det(e)} \epsilon^{\lambda\rho\gamma\delta}\epsilon^{\mu\nu\alpha\beta} \;
T^A_{\mu\nu}\,e^A_{\rho}\, \bar{\nabla}_\lambda \bar{R}_{\alpha\beta,\gamma\delta}, &
L'_{16} &=& \epsilon^{\alpha\beta\gamma\delta}\;
T^A_{\mu\nu}\;e^{A\,[\mu}\;\bar{\nabla}^{\nu]} \bar{R}_{\alpha\beta,\gamma\delta}, \\
L'_{17} &=& \epsilon^{\lambda\rho\alpha\beta}\;
T^A_{\mu\nu}\;e_{\rho}^A\;\bar{\nabla}_{\lambda} \bar{R}_{\alpha\beta}^{~~\mu\nu}, &
L'_{18} &=& \epsilon^{\lambda\rho\alpha\beta}\;
T^A_{\mu\nu}\;e_{\rho}^A\;\bar{\nabla}_{\lambda} \bar{R}^{\mu\nu}_{~~\alpha\beta}, \\
L'_{19} &=& T^A_{\mu\nu}\;e^{A\,\lambda}\;g^{\alpha[\mu}\;\epsilon^{\nu]\beta\gamma\delta}\;
\bar{\nabla}_{\lambda} \bar{R}_{\alpha\beta,\gamma\delta}, &
L'_{20} &=& T^A_{\mu\nu}\;e^{A\,\alpha}\;\epsilon^{\beta\gamma\delta[\mu}\;\bar{\nabla}^{\nu]} \bar{R}_{\alpha\beta,\gamma\delta}.
\end{array}
$$
However, many of these invariants are zero or reduce to one another when one takes into account the additional symmetries of the standard Riemann tensor $\bar{R}$,
\vspace{-0.2cm}
\bea\la{symR1}
\bar{R}_{\alpha\beta,\mu\nu}&=&\bar{R}_{\mu\nu,\alpha\beta}, \\
\la{symR2}
\epsilon^{\alpha\beta\mu\rho}\bar{R}_{\alpha\beta,\mu\nu}&=&0\qquad {\rm or}\qquad
\bar{R}_{\alpha\beta,\mu\nu} + \bar{R}_{\beta\mu,\alpha\nu} + \bar{R}_{\mu\alpha,\beta\nu} = 0,
\eea
\vspace{-0.2cm}
as well as the Bianchi identity,
\vspace{-0.02cm}
\beq
\varepsilon^{\alpha\beta\lambda\rho} \bar{\nabla}_\lambda \bar{R}_{\alpha\beta,\mu\nu}\;=\;0\qquad {\rm or}\qquad
\bar{\nabla}_\lambda \bar{R}_{\alpha\beta,\mu\nu} + \bar{\nabla}_\alpha \bar{R}_{\beta\lambda,\mu\nu}+ \bar{\nabla}_\beta \bar{R}_{\lambda\alpha,\mu\nu}\; = \;0.
\la{Bianchi1}\eeq
\vspace{-0.2cm}
Contracting \Eq{Bianchi1} with the metric tensor one obtains the identities for the Ricci tensor and the curvature:
\vspace{-0.2cm}
\beq
\bar{\nabla}_\beta \bar{R}^{\alpha\beta,\mu\nu} \;=\; -2 \bar{\nabla}^{[\mu} \bar{R}^{\nu]\alpha}, \qquad
\partial^{\mu} \bar R \;=\; 2 \bar{\nabla}_\nu \bar{R}^{\mu\nu}.
\la{Bianchi2}\eeq
\vspace{-0.2cm}
It is also helpful to keep in mind that the covariant derivative of the metric tensor and of the combination  $(1/\det(e))\,\epsilon^{\lambda\rho\mu\nu}$ are zero.

We immediately find that $L'_1=L'_2$, $L'_3=L'_4$, $L'_8=L'_9$, $L'_{10}=L'_{11}$, $L'_{17}=L'_{18}$ and $L'_8=0$ because of \Eq{symR1}, $L'_{12}=L'_{14}=L'_{15}=0$ because of \Eq{symR2}, and $L'_9=L'_{10}=L'_{11}=L'_{13}=0$ because of \Eq{Bianchi1}. The invariants $L'_1$, $L'_{2}$ and $L'_{6}$ are proportional owing to \Eq{Bianchi2}.
The invariants $L'_{10}=L'_{11}=0$ owing to \Eqs{symR2}{Bianchi1}. Therefore, actually only four linear independent invariants are left:
\bea\n
\begin{array}{ccccc}
L''_1 &=& \det(e)\;T^A_{\mu\nu}\;e^{A\,[\mu}\;\partial^{\nu]} \bar{R}
& = & -\det(e)\;v^\lambda\;\partial_\lambda \bar{R}, \\
L''_2 &=& \epsilon^{\lambda\rho\mu\nu}\;T^A_{\mu\nu}\;e_{\rho}^A\;\partial_{\lambda} \bar{R}
& = & 4\det(e)\,a^\lambda\;\partial_\lambda \bar{R}, \\
L''_3 &=& \det(e)\;T^A_{\mu\nu}\;e_{\rho}^A\;\bar{\nabla}^{[\mu} \bar{R}^{\nu]\rho}
& = & \frac{1}{3}\,\det(e)\;v_\lambda\;\partial^\lambda \bar{R}
+ \frac{2}{3}\,\det(e)\;t_{\rho,\mu\nu}\;\bar{\nabla}^{[\mu} \bar{R}^{\nu]\rho}, \\
L''_4 &=& \epsilon^{\lambda\rho\mu\nu}\;T^A_{\mu\nu}\;e^{A\,\sigma}\;\bar{\nabla}_{\lambda} \bar{R}_{\rho\sigma}
& = & \frac{2}{3}\;\det(e)\;a^\lambda\;\partial_\lambda \bar{R}
+ \frac{2}{3}\;\epsilon^{\lambda\rho\mu\nu}\;t^{~~\sigma}_{\mu\nu}\;\bar{\nabla}_{\lambda} \bar{R}_{\rho\sigma}\,,
\end{array}
\eea
and no linear combination of these invariants is a full derivative. They can be recombined in a simpler way:
\bea\la{L1-4}
\begin{array}{cccccc}
L_1 &=& \det(e)\,v^\lambda\,\partial_\lambda \bar{R}, & \quad
L_2 &=& \det(e)\,a^\lambda\;\partial_\lambda \bar{R}, \\
L_3 &=& \det(e)\,t_{\rho,\mu\nu}\,\bar{\nabla}^{[\mu} \bar{R}^{\nu]\rho}, & \quad
L_4 &=& \epsilon^{\lambda\rho\mu\nu}\,t^{~~\sigma}_{\mu\nu}\,\bar{\nabla}_{\lambda} \bar{R}_{\rho\sigma}\,.
\end{array}
\eea

Since these invariants are linear in torsion, they are potential sources of torsion even in the absence of fermions, including the reduced torsion part, $t_{\mu\nu}^{~~\lambda}$.

We end up the derivative expansion here. We do not consider four derivative terms of the
type $(\bar\nabla T)^2$, $(\bar\nabla T)T^2$, $\bar R T^2$ and $T^4$ (there are many dozens of such terms) since they lead to even smaller corrections to the Einstein equation than
the $T^2$ terms listed in Section VI.A and considered in the next Section, see the estimate in Section II. However, all four derivative terms are, by dimension (which is 4), on equal footing from the point of view of the ultraviolet renormalization of the theory about curved background with generally nonzero torsion. Therefore, they should all be included, for example, in the ``asymptotic safety'' approach~\cite{arXiv:0908.1964,arXiv:1110.6389}.
 
In the logic of the effective Lagrangians, which we assume in this paper, the gravitational action is an infinite series in $\partial^2/M_P^2$ such that it makes no sense in studying
the stability against small runaway fluctuations from flat space--time in the concrete $p^4$ order since the inverse propagator of the fields is an infinite series in $p^2/M_P^2$. What makes certain sense, is to study the stability of flat space--time at vanishing momenta but that is decided by the {\em two}-derivative terms. In addition to the usual condition that the Newton constant (or $M_P^2$) is positive, the new requirement is that the eigenvalues of the $T^2$ matrix are positive: this condition is summarized in the inequalities \ur{stability}.

It should be added that in the Einstein--Cartan formulation, none of the thinkable
diffeomorphism- and local Lorentz-invariant action terms is strictly speaking stable under large nonperturbative fluctuations of the frame and spin connection fields~\cite{arXiv:1109.0091}. This observation strengthens the argument that the present-day gravitation theory is but an effective low-energy one, and therefore it makes sense to study systematically order by order the possible effects of the higher derivative terms.

\section{Induced four-fermion interaction}

If we ignore the $T\nabla \bar R$ terms \ur{L1-4} that lead to high-order terms $(\nabla \bar R)^2$ if we exclude the torsion, we are left with terms quadratic in torsion \ur{bos-2} and terms linear in torsion coupled to fermion currents \ur{S-gamma-M}. It is important that in the leading order only the $a_\mu,v_\mu$ part of the torsion couples to fermions. In the next order, however, when a derivative is added, the reduced torsion part $t_{\mu\nu}^{~~\lambda}$ may also couple to fermions~\cite{Kostelecky:2007kx}.

In the leading order, if one integrates out the torsion the reduced torsion $t_{\mu\nu}^{~~\lambda}$ vanishes, whereas the Gaussian integral over $a_\mu,v_\mu$ produces the 4-fermion interaction Lagrangian
\bea\la{4ferm}
{\cal L}^{\Psi^4}&=&\frac{\det(e)}{2(M^2_{aa}M^2_{vv}-M^4_{av})}
\left[A_BA^B\(g_1^{+\;2}M^2_{vv}-2g_1^+g_2^+M^2_{av}+g_2^{+\;2}M^2_{aa}\)\right.\\
\n
&&\qquad\qquad\qquad\qquad +\,V_BV^B\(g_1^{-\;2}M^2_{vv}-2g_1^-g_2^-M^2_{av}+g_2^{-\;2}M^2_{aa}\) \\
\n
&&\qquad\qquad\qquad\qquad +\left.\,2A_BV^B\(g_1^+g_1^-M^2_{vv}-(g_1^+g_2^- + g_1^-g_2^+)M^2_{av}+g_2^+g_2^-M^2_{aa}\)\right]\\
\n
&\equiv&\sqrt{-g}\left(h_{AA}\,A_BA^B+h_{VV}\,V_BV^B+2h_{AV}\,A_BV^B\right).
\eea
where $A^B=\bar\Psi\gamma^B\gamma^5\Psi$ is the axial and $V^B=\bar\Psi\gamma^B\Psi$ is the vector current. The dimensionless constants $g^\pm_{1,2}$ are defined in \Eq{S-gamma-M} and the masses $M_{a,v}$ are defined in \Eq{bos-2}. The $A\cdot V$ interaction term is $C,P$-odd and $T$-even.

Certain particular cases of this Lagrangian have been considered before. For example, to
compare it with the paper by Freidel {\it et al.}~\cite{Freidel:2005sn} we take $g_1^+=-1$, $g_2^-=\alpha$ (see Section IV.C) and the `minimal model' values of the torsion masses \ur{MMmin} with the identification $\iota = -1/\gamma$, $M_P^2=2/G$. In this case our general \Eq{4ferm} reduces to Eq. (23) of Ref.~\cite{Freidel:2005sn}.

\section{Stress-energy tensor from four-fermion interaction}

If the $a_\mu, v_\mu$ masses are of the order of the Planck mass as in \Eq{MMmin} the 4-fermion Lagrangian \ur{4ferm} leads to a correction to the cosmological equation of the order of $p^2/M_P^2$ where $p$ is the characteristic momentum of the fermion matter, for example temperature. Therefore, it is a tiny correction unless $p$ approaches $M_P$ but then one has to take into account higher terms in the derivative expansion, that are being neglected.

As discussed in Section V, the addition of torsion to the General Relativity in the fermion sector enlarges the gauge symmetry of gravity from the Lorentz $SU(2)_{\rm L}\times SU(2)_{\rm R}$ group to the $U(2)_{\rm L}\times U(2)_{\rm R}$ group which we know must be broken by the spontaneous or explicit masses of the $a_\mu,v_\mu$ vector bosons. However, these masses need not be of the order of the Planck mass but could be much smaller, say, of the order of $10\;{\rm TeV}$~\cite{Belyaev:2007fn}. In this case the 4-fermion interaction \ur{4ferm} could be an important correction in the epoch preceding the electroweak phase transition.

Anyway, there is an interesting problem of evaluating the contribution of the 4-fermion interaction to the stress-energy tensor in the r.h.s. of the Einstein--Friedman cosmological equation. This problem has been addressed {\it e.g.} in Refs.~\cite{Nurgaliev:1983,Gasperini:1986mv,deBerredoPeixoto:2009zz,Poplawski:2010kb}
using the ideas of a ``spin fluid''~\cite{Weyssenhoff:1947,Ray:1982qs,Brechet:2007}. We think that this approach is unsatisfactory. Particles with spin one-half are always quantum, for example there are exchange effects, and that cannot be mimicked by any semi-classical model. At some point in the above references one has to average the spin-squared operator $<s^2>$. This quantity is replaced by $1/4$, why not $3/4$?

Meanwhile, averaging 4-fermion operators over a fermion medium is a common problem in Quantum Field Theory. With other fermion interactions (temporally) switched off, the contribution of the 4-fermion average to the Lagrangian is given by two terms -- the `direct' (Hartree) term and the `exchange' (Fock) term, corresponding to two possible contractions of the $\Psi,\bar\Psi$ operators by the fermion propagator $G(p)$, see Fig.~1:
\bea\la{HF1}
<(\bar\Psi\Gamma_1\Psi)(\bar\Psi\Gamma_2\Psi)> &=& \frac{1}{2}\int\!\frac{d^4p_1}{(2\pi)^4i}\,\Tr(G(p_1)\Gamma_1)
\int\!\frac{d^4p_2}{(2\pi)^4i}\,\Tr(G(p_2)\Gamma_2)\\
\n
&-&\frac{1}{2}\int\!\frac{d^4p_1}{(2\pi)^4i}\int\!\frac{d^4p_2}{(2\pi)^4i}\,\Tr(G(p_1)\Gamma_1G(p_2)\Gamma_2),
\eea
where $\Gamma_{1,2}$ can be arbitrary Dirac and fermion `flavor' matrices. If other fermion interactions need to be taken into account, one has to `dress' the propagators and the 4-vertex.
\vskip -1true cm

\begin{figure}[htb]
\begin{minipage}[]{.7\textwidth}
\includegraphics[width=\textwidth]{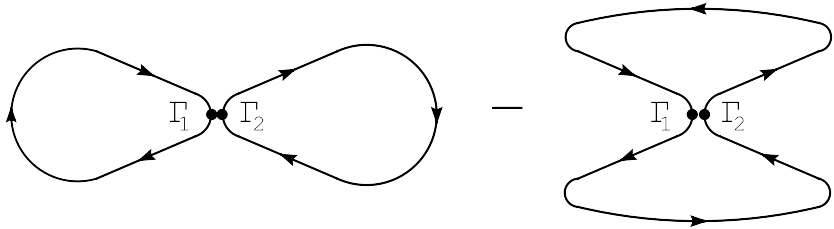}
\end{minipage}
\caption{Two contributions to the average of 4-fermion interaction: Hartree (left) and Fock (right).}
\label{fig:1}
\end{figure}

We emphasize that averaging over the medium should be performed in the Lagrangian; the corresponding correction to the stress-energy tensor is then obtained by varying the Lagrangian with respect to $g^{\mu\nu}$. Averaging in the equation of motion makes no sense.

In what follows we illustrate the use of \Eq{HF1} by taking a medium of non-interacting fermions.

\subsection{Averaging 4-fermion interaction in a non-interacting medium}

We consider one species of fermions with mass $m$ at temperature $T$ and chemical potential $\mu$ which
corresponds to certain charge density $\rho$, to be specified below. It should be stressed that in
a relativistic theory there is no strict way of separating particles from antiparticles: In a heat bath
both particles from the upper continuum and holes (antiparticles) from the lower continuum are excited;
the chemical potential regulates the {\it difference} between the number of particles and antiparticles,
which is the only well-defined quantity.

Neglecting interactions one writes the free fermion propagator
\beq
G(p)=\frac{1}{m-\Dirac{p}}=\frac{1}{m-(\mu+i\omega_n)\gamma^0-p_i\gamma^i}
=\frac{m+(\mu+i\omega_n)\gamma^0+p_i\gamma^i}{m^2+p^2-(\mu+i\omega_n)^2}
\la{prop}\eeq
where $\omega_n=2\pi T(n+\half)$ are the (imaginary) Matsubara frequencies. Integration over $p_0$ becomes a summation over Matsubara frequencies:
$$
\int\frac{dp_0}{2\pi i}\ldots = T\sum_{n=-\infty}^\infty\ldots
$$
The charge density of the fermion gas is given by
\bea\n
\rho &=& <j^0>=\langle\bar\Psi\gamma^0\Psi\rangle = T\sum_n \int\! \frac{d^3p}{(2\pi)^3}\,\Tr(G(p)\gamma^0)\\
&=& 4T\sum_n\int\!\frac{d^3p}{(2\pi)^3}\,\frac{\mu+i\omega_n}{\varepsilon^2-(\mu+i\omega_n)^2}
=2\int\!\frac{d^3p}{(2\pi)^3}\(\frac{1}{e^{\frac{\varepsilon-\mu}{T}}+1}-\frac{1}{e^{\frac{\varepsilon+\mu}{T}}+1}\),
\la{rho}\eea
where $\varepsilon=\sqrt{p^2+m^2}$. This is nothing but the difference between Fermi--Dirac distributions for
particles (positive $\mu$) and antiparticles (negative $\mu$). The integral can be easily evaluated in certain
limiting cases; in particular, in the ultra-relativistic case $m\ll\mu,T$ one has
\beq
\rho = \frac{\mu T^2}{3}+\frac{\mu^3}{3\pi^2}-m^2\,\frac{\mu}{2\pi^2}+{\cal O}(m^4).
\la{rho-ultra}\eeq
The second term dominates at high densities when the massless fermion gas becomes degenerate. One
can extract the chemical potential as function of $\rho$ from this equation.

The thermodynamic potential of the fermion gas is~\cite{landau:1980}
\bea\n
\Omega&=&-2VT\int\!\frac{d^3p}{(2\pi)^3}\log\left[\left(e^{\frac{\varepsilon-\mu}{T}}+1\right)
\left(e^{\frac{\varepsilon+\mu}{T}}+1\right)\right]\\
&=&-\left(\frac{7 \pi ^2}{180}\,T^4+\frac{1}{6}\,T^2 \mu ^2+\frac{1}{12 \pi ^2}\,\mu ^4\right)V+{\cal O}(m^2).
\la{Omega}\eea

We now compute the Hartree (direct) part of the 4-fermion interaction keeping in mind that in our case
$\Gamma_{1,2}=(\gamma^\mu,\gamma^\mu\gamma^5)$, see \Eq{4ferm}. We note immediately that the axial
current does not contribute to the Hartree part as $\Tr(G(p)\gamma^\mu\gamma^5)\equiv 0$. It could
have contributed were the chemical potential different for left- and right-handed particles but we
do not consider this possibility here. For the vector current we have two independent loop integrals
(and traces) each of which is exactly of the same type as for the calculation of the charge density.
Therefore, we obtain
\bea\la{H}
<V_BV^B>_{\rm Hart}&=&\langle(\bar\Psi\gamma_B\Psi)(\bar\Psi\gamma^B\Psi)\rangle_{\rm Hart}
=\frac{1}{2}\rho^2,\\
\n
<A_BA^B>_{\rm Hart}&=&0,\\
\n
<A_BV^B>_{\rm Hart}&=&0.
\eea

In the Fock (exchange) part the loop integrals and sums over Matsubara frequencies again factorize,
however the trace does not. We obtain
\bea\la{F}
<V_BV^B>_{\rm Fock}&=&\frac{1}{4}\rho^2-\frac{m^2}{2}\sigma^2,\\
\n
<A_BA^B>_{\rm Fock}&=&\frac{1}{4}\rho^2+\frac{m^2}{2}\sigma^2,\\
\n
<A_BV^B>_{\rm Fock}&=&0,
\eea
where we have denoted
\beq
\sigma= 4T\sum_n\int\!\frac{d^3p}{(2\pi)^3}\,\frac{1}{\varepsilon^2-(\mu+i\omega_n)^2}
=2\int\!\frac{d^3p}{\varepsilon(2\pi)^3}\[1-\frac{1}{e^{\frac{\varepsilon-\mu}{T}}+1}-\frac{1}{e^{\frac{\varepsilon+\mu}{T}}+1}\].
\la{sigma}\eeq
At $m\ll\mu,T$ one gets
$$
\sigma = -\frac{T^2}{6}-\frac{\mu^2}{2\pi^2}.
$$
The $\sigma^2$ terms combine with the ${\cal O}(m^2)$ corrections to the charge density \ur{rho-ultra}.

The full, Hartree plus Fock, contributions are
\bea\la{HF2}
<V_BV^B>_{\rm Hart+Fock}&=&\frac{3}{4}\rho^2-\frac{m^2}{2}\sigma^2,\\
\n
<A_BA^B>_{\rm Hart+Fock}&=&\frac{1}{4}\rho^2+\frac{m^2}{2}\sigma^2,\\
\n
<A_BV^B>_{\rm Hart+Fock}&=&0.
\eea
These expressions should be put in the 4-fermion Lagrangian \ur{4ferm} and we get:
\beq
{\cal L}^{\rm 4-ferm}=\sqrt{-g}\left[h_{VV}\left(\frac{3}{4}\rho^2-\frac{m^2}{2}\sigma^2\right)
+h_{AA}\left(\frac{1}{4}\rho^2+\frac{m^2}{2}\sigma^2\right)\right].
\la{4-ferm}\eeq
We stress again that $\rho$ here is the charge density, {\it i.e.} the density of particles {\it minus}
the density of antiparticles, therefore it is expected to be small in the cosmological context.
%
%It should be noted that, if fermions are interacting, the $<VV>$ part gets a correction linear
%in the interactions, whereas the $<AA>$ part differs from the free one only in the second
%order in interactions.
%\vspace{-0.5cm}

\subsection{Derivation of the stress-energy tensor from the Lagrangian}

If the Lagrangian is known, the corresponding stress-energy tensor if found by the general rule
\beq
\Theta_{\mu\nu}=2\,\frac{1}{\sqrt{-g}}\,\frac{\partial{\cal L}}{\partial g^{\mu\nu}}\quad+\quad
{\rm derivative\;terms}.
\la{Theta1}\eeq
The problem therefore is to establish the dependence of \ur{4-ferm} on the metric tensor. This can be
easily done if one realizes that $T$ and $\mu$ are actually the zero components of 4-vectors, as seen
from their use in \Eq{prop}.

Let us illustrate how this logic works by finding the correct stress-energy tensor in the leading
term in the fermion action. We take the first term in the thermodynamic potential \ur{Omega}, the simple
case of the Stefan--Boltzmann law for massless fermions, $\Omega=-7\pi^2T^4/180$. The partition function is
$$
{\cal Z}=\exp\left(-\frac{\Omega}{T}\right)=\exp\left(i\int\!dt\int\!d^3x(-\Omega/V)\right)
$$
where $t$ is now the Minkowski time. It means that the corresponding Minkowski Lagrangian ${\cal L}=-\Omega/V$
generalized to the case of an arbitrary metric is
$$
{\cal L}=\frac{7\pi^2}{180}\,\sqrt{-g}\,(T_\mu T_\nu g^{\mu\nu})^2,\qquad T_\mu = (T,0,0,0).
$$
Using the general \Eq{Theta1} we find immediately that in the co-moving frame
$$
\Theta_{\mu\nu}=\frac{7\pi^2}{180}T^4\left(-g_{\mu\nu}+4\,\delta_{\mu 0}\,\delta_{\nu 0}\right)
=\frac{7\pi^2}{60}\left\{\begin{array}{cc} T^4, & \mu=\nu=0,\\
\frac{1}{3}\,T^4, & \mu=\nu=1,2,3.\end{array}\right.\,,
$$
which gives, of course, the correct energy density $\epsilon=\Theta_{00}$ and pressure
$p=\Theta_{11}=\Theta_{22}=\Theta_{33}=\frac{\epsilon}{3}$ for the ultra-relativistic fermion gas.
In fact $\epsilon = 3p$ holds true for any relation between $T$ and $\mu$ in that gas, since
the dilatational current is conserved when there are no dimensional world constants, hence $\Theta^\mu_\mu=0$.

We now find the stress-energy tensor following from the 4-fermion Lagrangian \ur{4-ferm}. It will not satisfy
the relation $\epsilon=3p$ anymore since the couplings $h_{VV}$ and $h_{AA}$ are not dimensionless but are
of the order of $1/M^2$ where $M$ is the mass of $a_\mu,v_\mu$ bosons. Neglecting the fermion mass we have
\beq
{\cal L}^{\rm 4-ferm}=\left(\frac{3}{4}h_{VV}+\frac{1}{4}h_{AA}\right)\sqrt{-g}\,\left.\rho^2\right|_{m=0},\qquad
\left.\rho^2\right|_{m=0} = \frac{\mu^2 T^4}{9}+\frac{2\mu^4T^2}{9\pi^2}+\frac{\mu^6}{9\pi^4}.
\la{4-ferm-1}\eeq
At first glance, when we promote $\mu,T$ to be 4-vectors there is an ambiguity: one can write $\mu^2T^4$
as $(\mu\cdot\mu)(T\cdot T)^2$ or as $(\mu\cdot T)^2(T\cdot T)$ or their combination. However, the result
is independent of the decoding as long as $\mu$ and $T$ remain parallel, which is the case in the co-moving frame.
What counts, is the total power of the polynomial in $\mu,T$, which is 6 in this case. Using the
general rule \ur{Theta1} we obtain for ultra-relativistic fermions:
\beq
\left.\Theta^{\rm 4-ferm}_{\mu\nu}\right|_{m=0}
=\left(\frac{3}{4}h_{VV}+\frac{1}{4}h_{AA}\right)\left.\rho^2\right|_{m=0}
\left(-g_{\mu\nu} + 6\,\delta_{\mu 0}\,\delta_{\nu 0}\right)
\la{Theta2}\eeq
implying $\epsilon^{\rm 4-ferm}=5\,p^{\rm 4-ferm}$. This equation of state for the 4-fermion piece can be
independently checked by using the general thermodynamic relations, see {\it e.g.} Eq. (1.4) of
the book~\cite{kapusta:1989}. Indeed, if ${\cal Z}=\exp(-\Omega(\mu)/T)$ is the partition function one has
\bea\n
&&{\rm pressure}\;p=T\frac{\partial \ln{\cal Z}}{\partial V},\quad
{\rm charge}\;Q=T\frac{\partial \ln{\cal Z}}{\partial \mu},\quad
{\rm charge\;density}\;\rho=\frac{Q}{V},
\\
\n
&&{\rm entropy}\;S=\frac{\partial (T\ln{\cal Z})}{\partial T},\quad
{\rm energy}\;E=-pV+TS+\mu Q,\quad
{\rm energy\;density}\;\epsilon=\frac{E}{V}.
\eea
In our case we have from \Eq{4-ferm-1}
$$
\ln{\cal Z}^{\rm 4-ferm}=\left(\frac{3}{4}h_{VV}+\frac{1}{4}h_{AA}\right)\left.\rho^2\right|_{m=0}\frac{V}{T}.
$$
From the above general relations one immediately finds
$$
\epsilon^{\rm 4-ferm}(\mu,T) =5\left(\frac{3}{4}h_{VV}+\frac{1}{4}h_{AA}\right)\left.\rho^2\right|_{m=0}= 5\,p^{\rm 4-ferm}(\mu,T)
$$
confirming \Eq{Theta2}. The equation implies that $\mu$ and $T$ are used as independent variables.

However one may wish to express the stress-energy tensor in terms of the charge density $\rho$.
To that end, one has to solve the equation $\rho= (T/V)(\partial \ln{\cal Z}/\partial \mu)$ with respect to $\mu$
and to substitute the function $\mu(\rho)$ into the stress-energy tensor. In general, $\ln{\cal Z}$ is a sum
of the main piece \ur{Omega} and the 4-fermion piece \ur{4-ferm-1}, such that $\mu(\rho)$ is a complicated function.
But if the 4-fermion piece is a small perturbation, one can easily find the linear correction to the zero-order
$\mu(\rho)$ following from \Eq{rho-ultra}. In this case one obtains
$$
p^{\rm 4-ferm}(\rho,T)=-\left(\frac{3}{4}h_{VV}+\frac{1}{4}h_{AA}\right)
\left.\frac{\partial(\rho^2 V)}{\partial V}\right|_{Q={\rm const.}}
=\left(\frac{3}{4}h_{VV}+\frac{1}{4}h_{AA}\right)\rho^2=\epsilon^{\rm 4-ferm}(\rho,T).
$$

The relation $\epsilon^{\rm 4-ferm}=p^{\rm 4-ferm}$ has been used in
Refs.~\cite{Gasperini:1986mv,deBerredoPeixoto:2009zz,Poplawski:2010kb} as following from the ``spin fluid''
approach. We see, however, that it is valid only if the $\sigma$ term \ur{F} from the Fock exchange contribution
is neglected, and if the 4-fermion term is a small perturbation to the main part of the stress-energy tensor.
If the 4-fermion interaction is the leading term, as assumed for the early cosmological evolution in
the above references, the equation of state becomes $\epsilon = 5p$.
%\vspace{-0.5cm}

\subsection{Can the 4-fermion interaction be observable?}

On the whole, we come to rather pessimistic conclusions with regard to the observability of the possible
4-fermion interaction induced by integrating out torsion. If the 4-fermion constants $h_{VV},h_{AA}$ are
of the order of $1/M_P^2$ as assumed in the gravitational approach like in the `minimal model' discussed
(and criticized) above, it seems hopeless since it becomes significant only at particle momenta $p\sim M_P$
but at these momenta we do not know the theory at all and anyway the derivative expansion fails. In addition,
if for some reasons the 4-fermion interaction becomes large, it must be included into the equation of state
and not treated as a perturbation~\cite{Gasperini:1986mv,deBerredoPeixoto:2009zz,Poplawski:2010kb}
when it is overwhelming.

We have mentioned that the masses $M$ of $a_\mu,v_\mu$ bosons could be of non-gravitational origin and therefore
be, say, of the order of $10\;{\rm TeV}$. That would increase the torsion-induced 4-fermion interaction
by 30 orders of magnitude as compared to the previous case. Nevertheless, it is still hardly observable. In an
ultra-relativistic medium the correction of the 4-fermion interaction to the stress-energy tensor is of
the order of $(\mu^2/M^2)T^4$, as compared to the main contribution $\sim T^4$, see \Eq{4-ferm-1}. A chemical
potential in the TeV range is hardly imaginable.

In general, it makes sense to introduce the chemical potential only for conserved quantum numbers. There are many conserved
quantities in the late epoch, such as quark flavors and baryon (B) and lepton (L) numbers. However, in a late epoch
the density is small and the torsion effects are probably negligible. The earlier we go into the evolution the
fewer quantum numbers are conserved. It was thought some time ago that electroweak interactions break $B+L$ but
preserve $B-L$ quantum numbers, but today it is believed that both are broken since the Majorana type of neutrino
is preferable. It looks like there are no conserved numbers at all in the epoch preceding the electroweak phase
transition~\cite{Shaposhnikov:2009zz}, save the electric charge for which the chemical potential is zero.
Therefore, it may well happen that in that early epoch the only possible contribution to the 4-fermion interaction
is due not to charge density $\rho$ (which is zero) but to the Fock `exchange' part having the $m^2$ piece, where
$m$ is the fermion mass, see \Eq{F}. It may become large if there are super-heavy fermions but then their contribution
is suppressed by the Boltzmann factor $\exp(-m/T)$, unless the temperature is of the same order of magnitude.

Finally, we should mention the possibility that there is no thermal equilibrium in the epoch preceding the
electroweak transition, meaning that temperature is not an adequate quantity. The Matsubara propagator \ur{prop}
is then irrelevant and should be replaced by
$$
G(p)=\frac{1}{m-\frac{i}{2\tau}-\Dirac{p}}
$$
where $\tau$ is the relaxation time. In this case $1/\tau$ replaces, qualitatively, $T$ and $\mu$ in the above
equations for the estimate of the average 4-fermion interaction which, in principle, can then become sizable.
We also mention an interesting possibility that an interplay of the evolution out of thermal equilibrium, and
the potential $C,P$ violation by torsion may lead to the baryon asymmetry of the Universe.
%\vspace{-0.4cm}

\section{Conclusions}

We have systematically listed all possible invariants that may arise in General Relativity when one includes
torsion, following the guiding principle of the derivative expansion. These include all possible invariants
quadratic in torsion (5 invariants), quadratic in curvature (10), and linear in torsion and linear
in the covariant derivatives of curvature (4). In the fermion sector, we have derived four possible
invariants with torsion coupled to the bilinear fermion currents. We do not limit ourselves to $P,T$-even
invariants. Some of the invariants are new, although most of them have been considered by different people before.

In the leading one-derivative order, only 8 components of torsion (out of the general 24) couple to fermions,
which can be cast into the Abelian axial ($a_\mu$) and vector ($v_\mu$) fields. If parity is conserved,
the interaction of $a_\mu,v_\mu$ fields with fermions possesses gauge $U(1)_{\rm L}\times U(1)_{\rm R}$
symmetry, in addition to the Lorentz gauge symmetry $SU(2)_{\rm L}\times SU(2)_{\rm R}$. Linear combinations
of $a_\mu,v_\mu$ are the gauge bosons of this additional symmetry.

However, the bosonic torsion-squared invariants break explicitly this symmetry as they provide masses to the
Abelian bosons $a_\mu,v_\mu$ or their linear combinations. From this point of view, such mass terms
may look unnatural: spontaneous breaking could be more aesthetic. Certain justification for writing
terms quadratic in torsion comes from the fact that they appear anyway from expanding the Einstein--Cartan
action. This is called the `minimal model' for torsion. However, we have shown that the minimal model leads
to a non-positive mass matrix for the $a_\mu,v_\mu$ bosons, therefore the minimal model cannot be complete.

Assuming on purely phenomenological grounds that there is a positive mass matrix and neglecting
higher-derivative invariants, we integrate out the torsion field and obtain the effective four-fermion action.
It contains, generally speaking, axial-axial, axial-vector and vector-vector interactions. The effect of the first
one has been studied in the past, with regard to its application to the Einstein--Friedman cosmological equation,
using the so-called ``spin fluid'' approach. We find this approach unsound since particles with spin one-half
are always quantum (for example there are exchange effects) and that cannot be mimicked by any semi-classical model.
We present a systematic quantum field-theoretic method to average the 4-fermion interaction over the fermion
medium, and perform the explicit averaging in the case of free fermions with given chemical potential
and temperature. The result is essentially different from that of the ``spin fluid'' approach.

We arrive to rather pessimistic conclusions concerning the possibility to observe any effects of the torsion-induced
4-fermion interaction. However under certain circumstances it may have cosmological consequences, see Section VIII.C,
but this has not been worked out.
\vspace{-0.4cm}

\section*{Acknowledgements}

We thank Profs. Victor Petrov and Maxim Polyakov for helpful discussions and Profs. Friedrich Hehl and Ilya Shapiro
for correspondence. This work has been supported in part by Russian Government grants RFBR-06-02-16786 and
RSGSS-3628.2008.2, and by Deutsche Forschungsgemeinschaft (DFG) grant 436 RUS 113/998/01. A.T. acknowledges
a stipend by the Dynasty Foundation.
\vskip 0.3true cm


\begin{thebibliography}{99}

%\cite{Hehl:1976kj}
\bibitem{Hehl:1976kj}
  F.~W.~Hehl, P.~Von Der Heyde, G.~D.~Kerlick and J.~M.~Nester,
  %``General Relativity With Spin And Torsion: Foundations And Prospects,''
  Rev.\ Mod.\ Phys.\  {\bf 48}, 393 (1976).
  %%CITATION = RMPHA,48,393;%%

%\cite{Sezgin:1979zf}
\bibitem{Sezgin:1979zf}
  E.~Sezgin and P.~van Nieuwenhuizen,
  %``New Ghost Free Gravity Lagrangians With Propagating Torsion,''
  Phys.\ Rev.\  D {\bf 21}, 3269 (1980).
  %%CITATION = PHRVA,D21,3269;%%

%\cite{Hehl:1994ue}
\bibitem{Hehl:1994ue}
  F.~W.~Hehl, J.~D.~McCrea, E.~W.~Mielke and Y.~Ne'eman,
  %``Metric affine gauge theory of gravity: Field equations, Noether identities,
  %world spinors, and breaking of dilation invariance,''
  Phys.\ Rept.\  {\bf 258}, 1 (1995)
  [arXiv:gr-qc/9402012].
  %%CITATION = PRPLC,258,1;%%

%\cite{Shapiro:2001rz}
\bibitem{Shapiro:2001rz}
  I.~L.~Shapiro,
  %``Physical aspects of the space-time torsion,''
  Phys.\ Rept.\  {\bf 357}, 113 (2002).
  [hep-th/0103093].

%\cite{Hammond:2002rm}
\bibitem{Hammond:2002rm}
  R.~T.~Hammond,
  %``Torsion Gravity,''
  Rept.\ Prog.\ Phys.\  {\bf 65}, 599 (2002).
  %%CITATION = RPPHA,65,599;%%

%\cite{Khriplovich:2005jh}
\bibitem{Khriplovich:2005jh}
  I.~B.~Khriplovich, A.~A.~Pomeransky,
  %``Remark on Immirzi parameter, torsion, and discrete symmetries,''
  Phys.\ Rev.\  {\bf D73}, 107502 (2006)
  [arXiv:hep-th/0508136].

%\cite{Freidel:2005sn}
\bibitem{Freidel:2005sn}
  L.~Freidel, D.~Minic, T.~Takeuchi,
  %``Quantum gravity, torsion, parity violation and all that,''
  Phys.\ Rev.\  {\bf D72}, 104002 (2005)
  [arXiv:hep-th/0507253].

%\cite{Mielke:2006zp}
\bibitem{Mielke:2006zp}
  E.~W.~Mielke and E.~S.~Romero,
  %``Cosmological evolution of a torsion-induced quintaxion,''
  Phys.\ Rev.\  D {\bf 73}, 043521 (2006).
  %%CITATION = PHRVA,D73,043521;%%

%\cite{Zubkov:2010sx}
\bibitem{Zubkov:2010sx}
  M.~A.~Zubkov,
  %``Torsion instead of Technicolor,''
  Mod.\ Phys.\ Lett.\  {\bf A25}, 2885-2898 (2010).
  [arXiv:1003.5473 [hep-ph]].

%\cite{arXiv:0908.1964}
\bibitem{arXiv:0908.1964}
  S.~Weinberg,
  %``Effective Field Theory, Past and Future,''
  PoSCD\ {\bf 09}, 001  (2009)
  [arXiv:0908.1964 [hep-th]].
  %%CITATION = POSCI,CD09,001;%%

%\cite{Nurgaliev:1983}
\bibitem{Nurgaliev:1983}
  I.~S.~Nurgaliev, W.~N.~Ponomariev,
  %``The earliest evolutionary stages of the Universe and space-time torsion,''
  Phys.\ Lett. \ {\bf B130}, 378-379 (1983).

%\cite{Gasperini:1986mv}
\bibitem{Gasperini:1986mv}
  M.~Gasperini,
  %``SPIN DOMINATED INFLATION IN THE EINSTEIN-CARTAN THEORY,''
  Phys.\ Rev.\ Lett.\  {\bf 56}, 2873 (1986).
  %%CITATION = PRLTA,56,2873;%%

%\cite{deBerredoPeixoto:2009zz}
\bibitem{deBerredoPeixoto:2009zz}
  G.~de Berredo-Peixoto and E.~A.~De Freitas,
  %``On the cosmological effects of the Weyssenhoff spinning fluid in the
  %Einstein-Cartan framework,''
  Int.\ J.\ Mod.\ Phys.\  A {\bf 24}, 1652 (2009)
  [arXiv:0907.1701 [gr-qc]].
  %%CITATION = IMPAE,A24,1652;%%

%\cite{Poplawski:2010kb}
\bibitem{Poplawski:2010kb}
  N.~J.~Poplawski,
  %``Cosmology with torsion - an alternative to cosmic inflation,''
  Phys.\ Lett.\  {\bf B694}, 181-185 (2010)
  [arXiv:1007.0587 [astro-ph.CO]].

%\cite{Fock:1929vt}
\bibitem{Fock:1929vt}
V.~Fock and D.~Iwanenko,
% ``Linear Quantum Geometry and Parallel Transport,''
  C.\ R.\ Acad.\ Sci., Paris {\bf 188} (1929) 1470;
V.~Fock,
  % ``Geometrization of Dirac's theory of the electron,''
  Z.\ Phys.\ {\bf 57} (1929) 261;
  %%CITATION = ZEPYA,57,261;%%
V.~Fock,
  % ``On the Dirac Equations in General Relativity,''
  C.\ R.\ Acad.\ Sci., Paris {\bf 189} (1929) 25;
V.~Fock,
% ``Dirac Wave Equation and Riemann Geometry'',
Le Journal de Physique et de Radium, S\`erie VI, {\bf 10} (1929) 392.
[English translation of the last two papers in: V.A.~Fock, Selected works,
L.D.~Faddeev, L.A~Khalfin and I.V.~Komarov, eds., Chapman and Hall
% / CRC
(2004)].

%\cite{Weyl:1929fm}
\bibitem{Weyl:1929fm}
  H.~Weyl,
  % ``Electron and Gravitation I (In German),''
  Z.\ Phys.\  {\bf 56} (1929) 330
  [Surveys High Energ.\ Phys.\  {\bf 5} (1986) 261].
  %%CITATION = SHEPD,5,261;%%
%\cite{Hojman:1980kv}

\bibitem{Hojman:1980kv}
  R.~Hojman, C.~Mukku and W.~A.~Sayed,
  %``PARITY VIOLATION IN METRIC TORSION THEORIES OF GRAVITATION,''
  Phys.\ Rev.\  D {\bf 22} (1980) 1915.
  %%CITATION = PHRVA,D22,1915;%%

%\cite{Nelson:1980ph}
\bibitem{Nelson:1980ph}
  P.~C.~Nelson,
  %``GRAVITY WITH PROPAGATING PSEUDOSCALAR TORSION,''
  Phys.\ Lett.\  A {\bf 79}, 285 (1980).
  %%CITATION = PHLTA,A79,285;%%

%\cite{Holst:1995pc}
\bibitem{Holst:1995pc}
  S.~Holst,
  %``Barbero's Hamiltonian derived from a generalized Hilbert-Palatini action,''
  Phys.\ Rev.\  D {\bf 53}, 5966 (1996)
  [arXiv:gr-qc/9511026].
  %%CITATION = PHRVA,D53,5966;%%

%\cite{Tsamparlis:1981xm}
\bibitem{Tsamparlis:1981xm}
  M.~Tsamparlis,
  %``METHODS FOR DERIVING SOLUTIONS IN GENERALIZED THEORIES OF GRAVITATION: THE
  %EINSTEIN-CARTAN THEORY,''
  Phys.\ Rev.\  D {\bf 24}, 1451 (1981), Appendix A.
  %%CITATION = PHRVA,D24,1451;%%

%\cite{Capozziello:2001mq}
\bibitem{Capozziello:2001mq}
  S.~Capozziello, G.~Lambiase and C.~Stornaiolo,
  %``Geometric classification of the torsion tensor in space-time,''
  Annalen Phys.\  {\bf 10}, 713 (2001)
  [arXiv:gr-qc/0101038]. The notations in \Eq{T-split} are in fact varying from author to author.
  For reasons that will become clear below, we introduce in \Eq{T-split} new notations and
  normalizations for the $a_\mu, v_\mu$ and $t_{\mu\nu}^{~~\lambda}$ fields.
  %%CITATION = ANPYA,10,713;%%

%\cite{Perez:2005pm}
\bibitem{Perez:2005pm}
  A.~Perez and C.~Rovelli,
  %``Physical effects of the Immirzi parameter,''
  Phys.\ Rev.\  D {\bf 73}, 044013 (2006)
  [arXiv:gr-qc/0505081].
  %%CITATION = PHRVA,D73,044013;%%

%\cite{Kostelecky:2007kx}
\bibitem{Kostelecky:2007kx}
  V.~A.~Kostelecky, N.~Russell and J.~Tasson,
  %``New Constraints on Torsion from Lorentz Violation,''
  Phys.\ Rev.\ Lett.\  {\bf 100}, 111102 (2008)
  [arXiv:0712.4393 [gr-qc]].
  %%CITATION = PRLTA,100,111102;%%

%\cite{Vainshtein:1981wh}
\bibitem{Vainshtein:1981wh}
  A.~I.~Vainshtein, V.~I.~Zakharov, V.~A.~Novikov and M.~A.~Shifman,
  %``ABC's of Instantons,''
  Sov.\ Phys.\ Usp.\  {\bf 25}, 195 (1982)
  [Usp.\ Fiz.\ Nauk {\bf 136}, 553 (1982)].
  %%CITATION = UFNAA,136,553;%%

%\cite{deBerredoPeixoto:1999vj}
\bibitem{deBerredoPeixoto:1999vj}
  G.~de Berredo-Peixoto, J.~A.~Helayel-Neto and I.~L.~Shapiro,
  %``On the consistency of a fermion torsion effective theory,''
  JHEP {\bf 0002}, 003 (2000)
  [arXiv:hep-th/9910168].
  %%CITATION = JHEPA,0002,003;%%

%\cite{Belyaev:2007fn}
\bibitem{Belyaev:2007fn}
  A.~S.~Belyaev, I.~L.~Shapiro and M.~A.~B.~do Vale,
  %``Torsion phenomenology at the LHC,''
  Phys.\ Rev.\  D {\bf 75}, 034014 (2007)
  [arXiv:hep-ph/0701002].
  %%CITATION = PHRVA,D75,034014;%%

%\cite{Baekler:2010fr}
\bibitem{Baekler:2010fr}
  P.~Baekler, F.~W.~Hehl and J.~M.~Nester,
  %``Poincare gauge theory of gravity: Friedman cosmology with even and odd
  %parity modes. Analytic part,''
  Phys.\ Rev.\  D {\bf 83}, 024001 (2011)
  [arXiv:1009.5112 [gr-qc]]. We thank Prof. F.~Hehl for bringing our attention to this paper.
  %%CITATION = PHRVA,D83,024001;%%

%\cite{Nieh:1981ww}
\bibitem{Nieh:1981ww}
  H.~T.~Nieh and M.~L.~Yan,
  %``AN IDENTITY IN RIEMANN-CARTAN GEOMETRY,''
  J.\ Math.\ Phys.\  {\bf 23}, 373 (1982).
  %%CITATION = JMAPA,23,373;%%

%\cite{Mielke:2009zz}
\bibitem{Mielke:2009zz}
  E.~W.~Mielke,
  %``Topologically modified teleparallelism, passing through the Nieh-Yan
  %functional,''
  Phys.\ Rev.\  D {\bf 80}, 067502 (2009).
  %%CITATION = PHRVA,D80,067502;%%

%\cite{Mercuri:2006um}
\bibitem{Mercuri:2006um}
  S.~Mercuri,
  %``Fermions in Ashtekar-Barbero connections formalism for arbitrary values of
  %the Immirzi parameter,''
  Phys.\ Rev.\  D {\bf 73}, 084016 (2006)
  [arXiv:gr-qc/0601013].
  %%CITATION = PHRVA,D73,084016;%%

%\cite{Neville:1978bk}
\bibitem{Neville:1978bk}
  D.~E.~Neville,
  %``A Gravity Lagrangian With Ghost Free Curvature**2 Terms,''
  Phys.\ Rev.\  {\bf D18}, 3535 (1978).

%\cite{arXiv:1110.6389}
\bibitem{arXiv:1110.6389}
  R.~Percacci,
  %``A short introduction to asymptotic safety,''
  arXiv:1110.6389 [hep-th].
  %%CITATION = ARXIV:1110.6389;%%

%\cite{arXiv:1109.0091}
\bibitem{arXiv:1109.0091}
  D.~Diakonov,
  %``Towards lattice-regularized Quantum Gravity,''
  arXiv:1109.0091 [hep-th].
  %%CITATION = ARXIV:1109.0091;%%

%\cite{Weyssenhoff:1947}
\bibitem{Weyssenhoff:1947}
   J.~Weyssenhoff, A.~Raabe,
   %``Relativistic dynamics of spin-fluids and spin-particles,''
   Acta.\ Phys.\ Pol. {\bf 9}, 7 (1947).

%\cite{Ray:1982qs}
\bibitem{Ray:1982qs}
  J.~R.~Ray and L.~L.~Smalley,
  %``SPINNING FLUIDS IN GENERAL RELATIVITY,''
  Phys.\ Rev.\  D {\bf 26}, 2619 (1982);
  %%CITATION = PHRVA,D26,2619;%%
%\cite{Ray:1982qr}
%\bibitem{Ray:1982qr}
  %J.~R.~Ray and L.~L.~Smalley,
  %``SPINNING FLUIDS IN THE EINSTEIN-CARTAN THEORY,''
  %Phys.\ Rev.\
  {\it ibid.}
  D {\bf 27}, 1383 (1983).
  %%CITATION = PHRVA,D27,1383;%%

%\cite{Brechet:2007}
\bibitem{Brechet:2007}
  S.~D.~Brechet, M.~P.~Hobson, A.~N.~Lasenby,
  %``Weyssenhoff fluid dynamics in general relativity using a 1+3 covariant approach,''
  Class.\ Quant.\ Grav. {\bf 24}, 6329-6348 (2007).

%\cite{landau:1980}
\bibitem{landau:1980}
  L.~D.~Landau and E.~M.~Lifshitz,
  Statistical Physics, Third Edition, Part 1,
  Reed Educational and Professional Publishing, Oxford (1980), Section 61.

%\cite{kapusta:1989}
\bibitem{kapusta:1989}
  J.~I.~Kapusta,
  Finite-Temperature Field Theory,
  Cambridge University Press (1989).

%\cite{Shaposhnikov:2009zz}
\bibitem{Shaposhnikov:2009zz}
  M.~Shaposhnikov,
  %``Baryon asymmetry of the universe and neutrinos,''
  Prog.\ Theor.\ Phys.\  {\bf 122}, 185 (2009).
  %%CITATION = PTPKA,122,185;%%

\end{thebibliography}
\end{document}